\documentclass[12pt,a4paper]{article}
\makeatletter
\def\eqnarray{%
\stepcounter{equation}%
\let\@currentlabel=\theequation
\global\@eqnswtrue
\global\@eqcnt\z@
\tabskip\@centering
\let\\=\@eqncr
$$\halign to \displaywidth\bgroup\@eqnsel\hskip\@centering
$\displaystyle\tabskip\z@{##}$&\global\@eqcnt\@ne
\hfil$\displaystyle{{}##{}}$\hfil
&\global\@eqcnt\tw@$\displaystyle\tabskip\z@{##}$\hfil
\tabskip\@centering&\llap{##}\tabskip\z@\cr}
\makeatother
\newcommand{\ket}[1]{{\vert{#1}\rangle}}
\newcommand{\bra}[1]{{\langle{#1}\vert}}

\newcommand{\braket}[2]{{\langle{#1}\vert{#2}\rangle}}

\newcommand{\fukuso}{{\mathbf C}}
\newcommand{\real}{{\mathbf R}}
\newcommand{\futon}{{\bf N}}
\newcommand{\seisu}{{\bf Z}}

\newcommand{\zetta}[1]{{\vert{#1}\vert}}

\begin{document}

\title{\sl Standard and Non--Standard Quantum Models : \\
A Non--Commutative Version of the Classical System of SU(2) and 
SU(1,1) Arising from Quantum Optics}
\author{
  Kazuyuki FUJII
  \thanks{E-mail address : fujii@yokohama-cu.ac.jp}\\
  Department of Mathematical Sciences\\
  Yokohama City University\\
  Yokohama, 236--0027\\
  Japan
  }
\date{}
\maketitle
%
%
%
%
\begin{abstract}
  This is a challenging paper including some review and new results. 
  
  Since the non--commutative version of the classical system based on the 
  compact group $SU(2)$ has been constructed in (quant-ph/0502174) by making 
  use of Jaynes--Commings model and so--called Quantum Diagonalization Method 
  in (quant-ph/0502147), we construct a non--commutative version of the 
  classical system based on the non--compact group $SU(1,1)$ by modifying 
  the compact case. 
  
  In this model the Hamiltonian is not hermite but pseudo hermite, which 
  causes a big difference between two models. For example, in the classical 
  representation theory of $SU(1,1)$, unitary representations are 
  infinite dimensional from the starting point. 
  Therefore, to develop a unitary theory of non--commutative system of 
  $SU(1,1)$ we need an infinite number of non--commutative systems, 
  which means a kind of {\bf second non--commutativization}. This is a very 
  hard and interesting problem.
  
  We develop a corresponding theory though it is not always enough, and 
  present some challenging problems concerning how classical properties can 
  be extended to the non--commutative case. 
  
  This paper is arranged for the convenience of readers as the first 
  subsection is based on the standard model ($SU(2)$ system) and the next one 
  is based on the non--standard model ($SU(1,1)$ system). 
  This contrast may make the similarity and difference between the standard 
  and non--standard models clear.
\end{abstract}
%


%
%
%
%

\section{Introduction}

This is a challenging paper including some review of \cite{KF} and new 
results, and our ultimate aim is to construct a unified theory of 
Non--Commutative (Differential) Geometry and Quantum Computation.

The Hopf bundles (which are famous examples of fiber bundles) over 
${\bf K}={\bf R}$, ${\bf C}$, ${\bf H}$ (the field of quaternion numbers), 
${\bf O}$ (the field of octanion numbers) are classical objects and 
they are never written down in a local manner. If we write them locally then 
we are forced to encounter singular lines called the Dirac strings, 
see \cite{KF}, \cite{MN}. 

It is very interesting to comment that the Hopf bundles correspond to 
topological solitons called Kink, Monopole, Instanton, Generalized Instanton 
respectively, see for example \cite{MN}, \cite{Ra}, \cite{KFu}. 
Therefore they are very important objects to study. 

Berry has given another expression to the Hopf bundle and Dirac strings by 
making use of a Hamiltonian (a simple spin model including the parameters $x$, 
$y$ and $z$), see the paper(s) in \cite{SW}. 
We call this the Berry model for simplicity. In this paper let us restrict 
to the case of ${\bf K}$=${\bf C}$. 

We also construct a pseudo Berry model by replacing the Pauli matrices 
(the generators of $su(2)$) in the Hamiltonian with the generators of 
$su(1,1)$. For this model the ``Hamiltonian" is not hermite and a bundle is 
defined, which is called the pseudo Hopf bundle for simplicity. 
However, it is topologically trivial (therefore, there are no Dirac strings).

We would like to make the Hopf and pseudo Hopf bundles non--commutative. 
Whether such a generalization is meaningful or not is not clear at the 
current time, however it is worth trying, see for example \cite{AC}, 
\cite{BaI} or more recently \cite{SG} and its references.

By the way, we are studying a quantum computation based on Cavity QED and 
one of the basic tools is the Jaynes--Cummings model (or more generally 
the Tavis--Cummings one), \cite{JC}, \cite{MS}, \cite{papers}, \cite{FHKW}. 
This is given as a ``half" of the Dicke model under the resonance condition 
and rotating wave approximation associated to it. 
If the resonance condition is not taken, then this model gives a 
non--commutative version of the Berry model. However, this new one is 
different from usual one because $x$ and $y$ coordinates are quantized, 
while $z$ coordinate is not. 

We also construct a non--commutative version of the pseudo Berry model by 
replacing the generators as in the classical case. In this case, since 
the eigenvalues of the pseudo Hamiltonian should be real, the domain is 
extremely limited in the Fock space.

From the non--commutative Berry model we construct a non--commutative version 
of the Hopf bundle by making use of so--called Quantum Diagonalization Method 
developed in \cite{FHKSW}. 
Then we see that the Dirac strings appear in only states containing the ground 
one (${\cal F}\times \{\ket{0}\} \cup \{\ket{0}\}\times {\cal F}$), while they 
don't appear in excited states (${\cal F}\times {\cal F} 
- {\cal F}\times \{\ket{0}\} \cup \{\ket{0}\}\times {\cal F}$), where 
${\cal F}$ is the Fock space generated by 
$\{a,\ a^{\dagger},\ N=a^{\dagger}a\}$,

This means that classical singularities are not universal in the 
process of non--commutativization, which is a very interesting phenomenon. 
This is one of reasons why we consider non--commutative generalizations 
(which are not necessarily unique) of classical geometry.

We also construct a non--commutative version of the pseudo Hopf bundle in 
the non--commutative pseudo Berry model. Since in this case the bundle is 
trivial and there are no Dirac strings, the situation becomes easy.

Moreover, we construct a non--commutative version of the Veronese mapping 
which is the mapping from $\fukuso P^{1}$ to $\fukuso P^{n}$ with mapping 
degree $n$. The mapping degree is usually defined by making use of the 
(first--) Chern class, so our mapping will become important 
if a non--commutative (or quantum) ``Chern class" would be constructed.

We also construct a non--commutative version of the pseudo Veronese mapping 
which is the mapping from $\fukuso Q^{1}$ to $\fukuso Q^{n}$ with mapping 
degree $n$. 

We challenge to construct a non--commutative version of the spin 
representation of group $SU(2)$. However, our trial is not enough because 
we could not construct the general case except for the special cases of spin 
$j=1$ and $j=3/2$. 
In this problem, we meet a difficulty coming from the non--commutativity. 
Further study constructing a general theory will be required. 

We also challenge to construct a non--commutative version of the spin 
representation of group $SU(1,1)$. However, unitary representations are 
infinite dimensional from the starting point even in the classical case. 
To develop a unitary theory of non--commutative system of $SU(1,1)$ 
we need an infinite number of non--commutative systems, which means a kind of 
second non--commutativization. Therefore our trial is not enough, so that 
further study will be required. 

Why do we consider non--commutative versions of classical field models ? 
What is an advantage to consider such a generalization ? 
Such natural questions arise. This paper may give one of answers. 
Moreover, readers will find many interesting (challenging) problems. 

For the convenience of readers this paper is arranged as the first 
subsection is the system based on $SU(2)$ and the next one is the system 
based on $SU(1,1)$. 
This contrast may make the similarity and difference between the standard 
and non--standard models clear. We also add many appendices to make the text 
clear.

The contents of the paper are as follows :
\begin{flushleft}
{\bf Section 1}\ \ Introduction \\
{\bf Section 2}\ \ Mathematical Preliminaries \\
\quad \quad \quad \ {\bf 2.1}\ \ Classical SU(2) System $\cdots$ 
Compact Case \\
\quad \quad \quad \ {\bf 2.2}\ \ Classical SU(1,1) System $\cdots$ 
Non-Compact Case \\
{\bf Section 3}\ \ Standard and Non-Standard Berry Models and Dirac Strings \\
\quad \quad \quad \ {\bf 3.1}\ \ Standard Berry Model and Dirac Strings \\
\quad \quad \quad \ {\bf 3.1}\ \ Non-Standard Berry Model \\
{\bf Section 4}\ \ Non-Commutative Models Arizing from the 
Jaynes-Cummings Model \\
\quad \quad \quad \ {\bf 4.1}\ \ Standard Quantum Model \\
\quad \quad \quad \ {\bf 4.1}\ \ Non-Standard Quantum Model \\
{\bf Section 5}\ \ Non-Commutative Hopf and Pseudo Hopf Bundles  \\
\quad \quad \quad \ {\bf 5.1}\ \ Non-Commutative Hopf Bundle \\
\quad \quad \quad \ {\bf 5.2}\ \ Non-Commutative Pseudo Hopf Bundle \\
{\bf Section 6}\ \ Non-Commutative Veronese and Pseudo Veronese Mappings \\
\quad \quad \quad \ {\bf 6.1}\ \ Non-Commutative Veronese Mapping \\
\quad \quad \quad \ {\bf 6.2}\ \ Non-Commutative Pseudo Veronese Mapping \\
{\bf Section 7}\ \ Non-Commutative Representation Theory \\
\quad \quad \quad \ {\bf 7.1}\ \ Non-Commutative Version of $SU(2)$ Case \\
\quad \quad \quad \ {\bf 7.2}\ \ Non-Commutative Version of $SU(1,1)$ Case \\
{\bf Section 8}\ \ Discussion \\
{\bf Appendix}\ \  \\
\quad \quad \quad \ \ {\bf A}\ \ Classical Theory of Projective Spaces \\
\quad \quad \quad \ \ {\bf B}\ \ Local Coordinate of the Projector \\
\quad \quad \quad \ \ {\bf C}\ \ Some Calculations of First Chern Class \\
\quad \quad \quad \ \ {\bf D}\ \ Difficulty of Tensor Decomposition \\
\quad \quad \quad \ \ {\bf E}\ \ Calculation of Some Integrals
\end{flushleft}

\section{Mathematical Preliminaries}

In this section we prepare some mathematical preliminaries for the 
following sections.

\subsection{Classical $SU(2)$ System $\cdots$ Compact Case}

The compact Lie group $SU(2)$ and its Lie algebra $isu(2)$ ($i=\sqrt{-1}$) are 
\begin{equation}
\label{eq:SU(2)}
SU(2)=\left\{A \in M(2;\fukuso)\ |\ A^{\dagger}A=1_{2},\quad 
\mbox{det}(A)=1\right\}
\end{equation}
and 
\begin{equation}
\label{eq:su(2)}
su(2)=\left\{X \in M(2;\fukuso)\ |\ X^{\dagger}=X,\quad \mbox{tr}(X)=0
\right\}.
\end{equation}

The algebra is generated by the famous Pauli matrices 
$\sigma_{j}\ (j=1\sim 3)$
\[
\sigma_{1}=
\left(
  \begin{array}{cc}
      & 1 \\
    1 & 
  \end{array}
\right),
\quad
\sigma_{2}=
\left(
  \begin{array}{cc}
      & -i \\
    i & 
  \end{array}
\right),
\quad 
\sigma_{3}=
\left(
  \begin{array}{cc}
    1 &    \\
      & -1
  \end{array}
\right);\quad
1_{2}=
\left(
  \begin{array}{cc}
    1 &   \\
      & 1
  \end{array}
\right),
\]
and the map $su(2)\ \longrightarrow\ SU(2)$ is given as 
\[
\sum_{j=1}^{3}x_{j}\sigma_{j}\ \longrightarrow\ 
\mbox{exp}\left(i\sum_{j=1}^{3}x_{j}\sigma_{j}\right).
\]

We usually use
\[
\sigma_{+}\equiv (1/2)(\sigma_{1}+i\sigma_{2})=
\left(
  \begin{array}{cc}
    0 & 1 \\
    0 & 0
  \end{array}
\right), \quad 
\sigma_{-}\equiv (1/2)(\sigma_{1}-i\sigma_{2})=
\left(
  \begin{array}{cc}
    0 & 0 \\
    1 & 0
  \end{array}
\right).
\]
Then the $su(2)$ relation 
\[
[\tilde{\sigma}_{3},\sigma_{+}]=\sigma_{+},\quad
[\tilde{\sigma}_{3},\sigma_{-}]=-\sigma_{-},\quad
[\sigma_{+},\sigma_{-}]=2\tilde{\sigma}_{3}
\]
is well--known, where $\tilde{\sigma}_{3}=(1/2)\sigma_{3}$.

Let us note that
\begin{equation}
A=
\left(
  \begin{array}{cc}
    \alpha & -\bar{\beta} \\
    \beta  &  \bar{\alpha}
  \end{array}
\right),\quad |\alpha|^{2}+|\beta|^{2}=1
\end{equation}
is an element in $SU(2)$.

\subsection{Classical $SU(1,1)$ System $\cdots$ Non--Compact Case}

The non--compact Lie group $SU(1,1)$ and its Lie algebra $isu(1,1)$ are 
\begin{equation}
\label{eq:SU(1,1)}
SU(1,1)=\left\{B \in M(2;\fukuso)\ |\ B^{\dagger}JB=J,\quad 
\mbox{det}(B)=1\right\}
\end{equation}
and 
\begin{equation}
\label{eq:su(1,1)}
su(1,1)=\left\{Y \in M(2;\fukuso)\ |\ Y^{\dagger}=JYJ,\quad \mbox{tr}(Y)=0
\right\}
\end{equation}
where $J=\sigma_{3}$. The algebra is generated by the matrices 
$\tau_{j}\ (j=1\sim 3)$
\[
\tau_{1}=
\left(
  \begin{array}{cc}
      & 1 \\
   -1 & 
  \end{array}
\right),
\quad
\tau_{2}=
\left(
  \begin{array}{cc}
      & -i \\
   -i & 
  \end{array}
\right),
\quad 
\tau_{3}=
\left(
  \begin{array}{cc}
    1 &   \\
      & -1
  \end{array}
\right)
=\sigma_{3},
\]
and the map $su(1,1)\ \longrightarrow\ SU(1,1)$ is given as 
\[
\sum_{j=1}^{3}x_{j}\tau_{j}\ \longrightarrow\ 
\mbox{exp}\left(i\sum_{j=1}^{3}x_{j}\tau_{j}\right).
\]

We usually use
\[
\tau_{+}\equiv (1/2)(\tau_{1}+i\tau_{2})=
\left(
  \begin{array}{cc}
    0 & 1 \\
    0 & 0
  \end{array}
\right), \quad 
\tau_{-}\equiv (1/2)(\tau_{1}-i\tau_{2})=
\left(
  \begin{array}{cc}
    0 & 0 \\
   -1 & 0
  \end{array}
\right).
\]
Then the $su(1,1)$ relation 
\[
[\tilde{\tau}_{3},\tau_{+}]=\tau_{+},\quad
[\tilde{\tau}_{3},\tau_{-}]=-\tau_{-},\quad
[\tau_{+},\tau_{-}]=-2\tilde{\tau}_{3}
\]
is well--known, where $\tilde{\tau}_{3}=(1/2)\tau_{3}$.

Let us note that
\begin{equation}
B=
\left(
  \begin{array}{cc}
    \alpha & -\bar{\beta} \\
   -\beta  &  \bar{\alpha}
  \end{array}
\right),\quad |\alpha|^{2}-|\beta|^{2}=1
\end{equation}
is an element in $SU(1,1)$.

\section{Standard and Non--Standard Berry Models and Dirac Strings}

We explain the way which Berry used in \cite{SW} to construct 
the Hopf bundle and Dirac strings corresponding to the compact case, 
and next construct ones corresponding to the non--compact case.

\subsection{Standard Berry Model and Dirac Strings}

The Hamiltonian used by Berry is a simple spin model 
\begin{equation}
\label{eq:berry-hamiltonian}
H_{B}
=x\sigma_{1}+y\sigma_{2}+z\sigma_{3}
=(x-iy)\sigma_{+}+(x+iy)\sigma_{-}+z\sigma_{3}
=
\left(
  \begin{array}{cc}
    z    & x-iy \\
    x+iy & -z
  \end{array}
\right)
\end{equation}
where $x$, $y$ and $z$ are parameters. This Hamiltonian is of course hermite. 
We would like to diagonalize $H_{B}$ above. 
The eigenvalues are 
\[
\lambda=\pm r\equiv \pm\sqrt{x^{2}+y^{2}+z^{2}}
\]
and corresponding orthonormal eigenvectors are 
\[
\ket{r}=\frac{1}{\sqrt{2r(r+z)}}
 \left(
  \begin{array}{c}
    r+z  \\
    x+iy   
  \end{array}
\right),\quad 
\ket{-r}=\frac{1}{\sqrt{2r(r+z)}}
 \left(
  \begin{array}{c}
    -x+iy  \\
    r+z      
  \end{array}
\right).
\]
Here we assume $(x,y,z) \in \real^{3}-\{(0,0,0)\}\equiv 
\real^{3}\setminus \{0\}$ to avoid a degenerate case. 
Therefore a unitary matrix defined by 
\begin{equation}
A_{I}=(\ket{r},\ket{-r})
=\frac{1}{\sqrt{2r(r+z)}}
\left(
  \begin{array}{cc}
    r+z  & -x+iy \\
    x+iy & r+z
  \end{array}
\right)
\end{equation}
makes $H_{B}$ diagonal like 
\begin{equation}
H_{B}=
A_{I}
\left(
  \begin{array}{cc}
    r &     \\
      & -r
  \end{array}
\right)
A_{I}^{\dagger}\equiv
A_{I}D_{B}A_{I}^{\dagger}.
\end{equation}
We note that the unitary matrix $A_{I}$ is not defined on the whole space 
$\real^{3}\setminus \{0\}$. The defining region of $U_{I}$ is 
\begin{equation}
D_{I}=\real^{3}\setminus \{0\}- \{(0,0,z)\in \real^{3}|\ z < 0\}.
\end{equation}
The removed line $\{(0,0,z)\in \real^{3}|\ z < 0\}$ 
is just the (lower) Dirac string, which is impossible to add to $D_{I}$.

Next, we have another diagonal form of $H_{B}$ like
\begin{equation}
H_{B}=A_{II}D_{B}A_{II}^{\dagger}
\end{equation}
with the unitary matrix $A_{II}$ defined by 
\begin{equation}
A_{II}
=\frac{1}{\sqrt{2r(r-z)}}
\left(
  \begin{array}{cc}
    x-iy & -r+z \\
    r-z  & x+iy
  \end{array}
\right).
\end{equation}
The defining region of $A_{II}$ is 
\begin{equation}
D_{II}=\real^{3}\setminus \{0\}- \{(0,0,z)\in \real^{3}|\ z > 0\}.
\end{equation}
The removed line $\{(0,0,z)\in \real^{3}|\ z > 0\}$ 
is the (upper) Dirac string, which is also impossible to add to $D_{II}$.

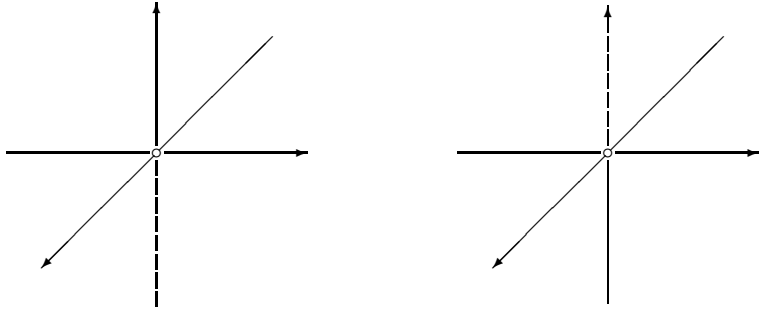
\begin{figure}
\begin{center}
\setlength{\unitlength}{1mm} 
\begin{picture}(140,60)(5,0)
\put(40,20){\circle{1}}
\put(20,20){\line(1,0){19}}
\put(41,20){\vector(1,0){19}}
\put(40,21){\vector(0,1){19}}
\multiput(40,19)(0,-2.5){8}{\line(0,-1){2}}
\put(40.4,20.4){\line(1,1){15}}
\put(39.7,19.7){\vector(-1,-1){15}}
\put(100,20){\circle{1}}
\put(80,20){\line(1,0){19}}
\put(101,20){\vector(1,0){19}}
\multiput(100,21)(0,2.5){7}{\line(0,1){2}}
\put(100,37.5){\vector(0,1){2}}
\put(100,19){\line(0,-1){19}}
\put(100.4,20.4){\line(1,1){15}}
\put(99.7,19.7){\vector(-1,-1){15}}
\end{picture}
\end{center}
\caption{Dirac strings corresponding to I and II}
\end{figure}

Here we have diagonalizations of two types for $H_{B}$, so a natural question 
comes about. What is a relation between $A_{I}$ and $A_{II}$ ? 
If we define 
\begin{equation}
\Phi
=\frac{1}{\sqrt{x^{2}+y^{2}}}
\left(
  \begin{array}{cc}
    x-iy &      \\
         & x+iy
  \end{array}
\right)
\end{equation}
then it is easy to see 
\[
A_{II}=A_{I}\Phi.
\]
We note that $\Phi$ (which is called a transition function) is not defined on 
the whole $z$--axis. 

What we would like to emphasize here is that the diagonalization of $H_{B}$ is 
not given globally (on $\real^{3}\setminus \{0\}$). However, the dynamics is 
perfectly controlled by the system 
\begin{equation}
\label{eq:Hopf-bundle}
\left\{(A_{I},D_{I}), (A_{II},D_{II}), \Phi, 
D_{I}\cup D_{II}=\real^{3}\setminus \{0\}\right\},
\end{equation}
which defines a famous fiber bundle called the Hopf bundle associated to the 
complex numbers ${\bf C}$ \footnote{The base space $\real^{3}\setminus \{0\}$ 
is homotopic to the two--dimensional sphere $S^{2}$},
\[
S^{1}\longrightarrow S^{3}\longrightarrow S^{2},
\]
see \cite{MN}.

The projector corresponding to the Hopf bundle is given as 
\begin{equation}
\label{eq:projector}
P(x,y,z)=A_{I}P_{0}A_{I}^{\dagger}=A_{II}P_{0}A_{II}^{\dagger}
=
\frac{1}{2r}
\left(
  \begin{array}{cc}
    r+z  & x-iy  \\
    x+iy & r-z
  \end{array}
\right),
\end{equation}
where $P_{0}$ is the basic one 
\[
P_{0}=
\left(
  \begin{array}{cc}
    1 &   \\
      & 0
  \end{array}
\right)\ \ {\in}
 \ \ M(2,{\bf C}).
\]
It is well--known that $P$ satisfies the relations
\[
1)\ P^{2}=P,\quad  2)\ P=P^{\dagger},\quad  3)\ \mbox{tr}P=1.
\]

We note that in (\ref{eq:projector}) Dirac strings don't appear because 
the projector $P$ is expressed globally.

\subsection{Non--Standard Berry Model}

The ``Hamiltonian" that we consider here is a modified one of the Berry model 
\begin{equation}
\label{eq:pseudo-berry-hamiltonian}
H_{pB}
=x\tau_{1}+y\tau_{2}+z\tau_{3}
=(x-iy)\tau_{+}+(x+iy)\tau_{-}+z\tau_{3}
=
\left(
  \begin{array}{cc}
    z      & x-iy \\
   -(x+iy) & -z
  \end{array}
\right)
\end{equation}
where $x$, $y$ and $z$ are parameters. This is not hermite. 
As a tentative terminology we call this a pseudo Berry model. 
We would like to diagonalize $H_{pB}$. 
The eigenvalues are 
\[
\lambda=\pm s\equiv \pm\sqrt{z^{2}-x^{2}-y^{2}},
\]
so the defining domain is 
\[
D\equiv 
\left\{(x,y,z) \in \real^{3}\ |\ z^{2}-x^{2}-y^{2} > 0\right\}
\]
Here, to avoid a degenerate case of eigenvalues we removed the case of 
$z^{2}-x^{2}-y^{2}=0$. We note that $D$ is not connected and consists of 
two domains $D_{+}$ and $D_{-}$ defined by 
\begin{equation}
D_{+}=\left\{(x,y,z) \in D\ |\ z > 0\right\}\quad \mbox{and}\quad 
D_{-}=\left\{(x,y,z) \in D\ |\ z < 0\right\}.
\end{equation}

The corresponding orthonormal eigenvectors are 
\[
\ket{s}=\frac{1}{\sqrt{2s(s+z)}}
 \left(
  \begin{array}{c}
    r+z    \\
   -(x+iy)   
  \end{array}
\right),\quad 
\ket{-s}=\frac{1}{\sqrt{2s(s+z)}}
 \left(
  \begin{array}{c}
    -x+iy  \\
     s+z      
  \end{array}
\right).
\]
Therefore a matrix defined by 
\begin{equation}
B_{I}=(\ket{s},\ket{-s})
=\frac{1}{\sqrt{2s(s+z)}}
\left(
  \begin{array}{cc}
    s+z   & -x+iy \\
  -(x+iy) & s+z
  \end{array}
\right)
\end{equation}
makes $H_{pB}$ diagonal like 
\begin{equation}
H_{pB}=
B_{I}
\left(
  \begin{array}{cc}
    s &     \\
      & -s
  \end{array}
\right)
B_{I}^{-1}\equiv
B_{I}D_{pB}B_{I}^{-1}.
\end{equation}
We note that the matrix $B_{I}$ is an element of the non--compact group 
$SU(1,1)$, and is not defined on $D_{-}$ because $s+z < 0$. Moreover, 
$B_{I}$ is defined on the whole $D_{I}=D_{+}$, so there is no singular line 
like Dirac strings.

\vspace{3mm}
{\bf A comment is in order}.\ We have another diagonal form of $H_{pB}$ like
\begin{equation}
H_{pB}=B_{II}D_{pB}B_{II}^{-1}
\end{equation}
with the matrix $B_{II}$ in $SU(1,1)$ defined by 
\begin{equation}
B_{II}
=\frac{1}{\sqrt{2s(z-s)}}
\left(
  \begin{array}{cc}
   -x+iy &  z-s    \\
    z-s  & -(x+iy)
  \end{array}
\right).
\end{equation}
The defining region of $B_{II}$ is 
\begin{equation}
D_{II}=D_{+}-\{(0,0,z)\in D_{+}\}.
\end{equation}
The removed line $\{(0,0,z)\in D_{+}\}$ 
is the (upper) Dirac string, which is also impossible to add to $D_{II}$.

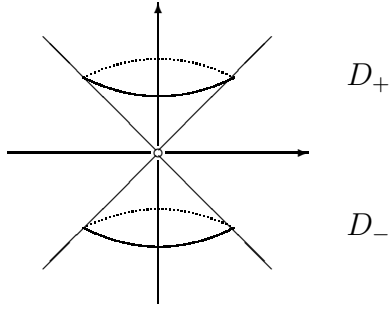
\begin{figure}
\begin{center}
\setlength{\unitlength}{1mm} 
\begin{picture}(80,60)(5,0)
\put(40,20){\circle{1}}
\put(20,20){\line(1,0){19}}
\put(41,20){\vector(1,0){19}}
\put(65,27){\makebox(6,6)[c]{$D_{+}$}}
\put(65, 7){\makebox(6,6)[c]{$D_{-}$}}
\put(40,19){\line(0,-1){19}}
\put(40,21){\vector(0,1){19}}
\put(40.1,20.4){\line(1,1){15}}
\put(39.7,19.6){\line(-1,-1){15}}
\put(39.7,20.4){\line(-1,1){15}}
\put(40.1,19.6){\line(1,-1){15}}
\bezier{200}(30,30)(40,25)(50,30)
\bezier{40}(30,30)(40,35)(50,30)
\bezier{200}(30,10)(40,5)(50,10)
\bezier{40}(30,10)(40,15)(50,10)
\end{picture}
\end{center}
\caption{The domains $D_{+}$ and $D_{-}$}
\end{figure}

However, with a singular transformation $\Phi$ (not defined on the whole 
$z$--axis) defined by
\begin{equation}
\Phi
=\frac{1}{\sqrt{x^{2}+y^{2}}}
\left(
  \begin{array}{cc}
    -x-iy &        \\
          & -x+iy
  \end{array}
\right)
\end{equation}
we can remove it because
\[
B_{I}=B_{II}\Phi.
\]

What we would like to emphasize in this case is that the diagonalization of 
$H_{pB}$ is given globally on $D_{I}$, which is very different from the 
compact case.
\begin{equation}
\label{eq:pseudo-Hopf-bundle}
\left\{H_{pB},B_{I},D_{I}\right\}.
\end{equation}
Here, as a tentative terminology we call this system a 
{\bf pseudo Hopf bundle} corresponding to the Hopf bundle in the compact case. 
However, this doesn't define a topological object because the domain $D_{I}$ 
is contractible (trivial in the sense of topology).

The projector corresponding to the case is given as 
\begin{equation}
\label{eq:pseudo-projector}
Q(x,y,z)=B_{I}Q_{0}B_{I}^{-1}
=
\frac{1}{2s}
\left(
  \begin{array}{cc}
     z+s   & x-iy  \\
   -(x+iy) & -z+s
  \end{array}
\right),
\end{equation}
where $Q_{0}=P_{0}$. 

$Q$ satisfies the relations
\[
1)\ Q^{2}=Q,\quad  2)\ JQJ=Q^{\dagger},\quad  3)\ \mbox{tr}Q=1.
\]

In the following we omit the suffix $I$ for simplicity.

\section{Non--Commutative Models Arising from the \\
Jaynes--Cummings Model}

In this section let us explain the Jaynes--Cummings model which is well--known 
in quantum optics, see \cite{JC}, \cite{MS}. From this we obtain a standard 
quantum model which is a natural extension of the (classical) Berry model. 
On the other hand, we obtain a non--standard quantum model by replacing 
the bases of $su(2)$ with the bases of $su(1,1)$.

\subsection{Standard Quantum Model}

The Hamiltonian of Jaynes--Cummings model can be written as follows (we set 
$\hbar=1$ for simplicity) 
\begin{equation}
\label{eq:hamiltonian}
H=
\omega 1_{2}\otimes a^{\dagger}a + \frac{\Delta}{2} \sigma_{3}\otimes {\bf 1} 
+ g\left(\sigma_{+}\otimes a+\sigma_{-}\otimes a^{\dagger} \right),
\end{equation}
where $\omega$ is the frequency of single radiation field, $\Delta$ the energy 
difference of two level atom, $a$ and $a^{\dagger}$ are annihilation and 
creation operators of the field, and $g$ a coupling constant. We assume that 
$g$ is small enough (a weak coupling regime). 
See the figure 3 as an image of the Jaynes--Cummings model (we don't repeat 
here). 

\begin{figure}
\begin{center}
\setlength{\unitlength}{1mm} 
\begin{picture}(80,40)(0,-20)
\bezier{200}(20,0)(10,10)(20,20)
\put(20,0){\line(0,1){20}}
\put(30,10){\circle*{3}}
\bezier{200}(40,0)(50,10)(40,20)
\put(40,0){\line(0,1){20}}
\put(10,10){\dashbox(40,0)}
\put(49,9){$>$}
\put(54,9){a photon}
\end{picture}
\end{center}
\vspace{-20mm}
\caption{One atom and a single photon inserted in a cavity}
\end{figure}
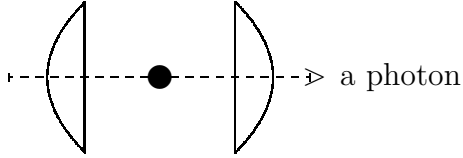

Now we consider the evolution operator of the model. 
We rewrite the Hamiltonian (\ref{eq:hamiltonian}) as follows. 
\begin{equation}
\label{eq:hamiltonian-rewrite}
H=
\omega 1_{2}\otimes a^{\dagger}a + \frac{\omega}{2}\sigma_{3}\otimes {\bf 1} + 
\frac{\Delta-\omega}{2} \sigma_{3}\otimes {\bf 1} +
g\left(\sigma_{+}\otimes a+\sigma_{-}\otimes a^{\dagger} \right)
\equiv H_{1}+H_{2}.
\end{equation}
Then it is easy to see $[H_{1},H_{2}]=0$, which leads to 
$
\mbox{e}^{-itH}=\mbox{e}^{-itH_{1}}\mbox{e}^{-itH_{2}}.
$

In the following we consider $\mbox{e}^{-itH_{2}}$ in which 
the resonance condition $\Delta-\omega=0$ is not taken. For simplicity 
we set $\theta=\frac{\Delta-\omega}{2g}(\neq 0)$ \footnote{Since the 
Jaynes--Cummings model is obtained by the Dicke model under some resonance 
condition on parameters included, it is nothing but an approximate one 
in the neighborhood of the point, so we must assume that $|\theta|$ is small 
enough. However, as a model in mathematical physics there is no problem 
to take $\theta$ be arbitrary} then 
\[
H_{2}
=g\left(
\sigma_{+}\otimes a+\sigma_{-}\otimes a^{\dagger} + 
\frac{\Delta-\omega}{2g}\sigma_{3}\otimes {\bf 1}
\right)
=g\left(
\sigma_{+}\otimes a+\sigma_{-}\otimes a^{\dagger} + 
\theta \sigma_{3}\otimes {\bf 1}
\right).
\]
For further simplicity we set 
\begin{equation}
\label{eq:non-commutative Berry model}
H_{JC}
=\sigma_{+}\otimes a+\sigma_{-}\otimes a^{\dagger} + 
\theta \sigma_{3}\otimes {\bf 1}
=
\left(
  \begin{array}{cc}
    \theta      & a       \\
    a^{\dagger} & -\theta
  \end{array}
\right),\quad [a,a^{\dagger}]={\bf 1}
\end{equation}
where we have written $\theta$ in place of $\theta {\bf 1}$ for simplicity. 

$H_{JC}$ can be considered as a non-commutative version of $H_{B}$ under the 
correspondence 
$
a\ \longleftrightarrow\ x-iy,\ a^{\dagger}\ \longleftrightarrow\ x+iy \ 
\mbox{and}\ \theta\ \longleftrightarrow\ z
$ : 
\begin{equation}
\label{eq:compact-type}
H_{B}
=
\left(
  \begin{array}{cc}
    z    & x-iy \\
    x+iy & -z
  \end{array}
\right),\ [x-iy,x+iy]=0 \ \longrightarrow \
H_{JC}
=
\left(
  \begin{array}{cc}
    \theta      & a       \\
    a^{\dagger} & -\theta
  \end{array}
\right),\ [a,a^{\dagger}]={\bf 1}.
\end{equation}
That is, $x$ and $y$ coordinates are quantized, while $z$ coordinate is not, 
which is different from usual one, see for example \cite{BaI}. 
It may be possible for us to call this {\bf a non--commutative Berry model}. 
We note that this model is derived not ``by hand" but by the model in quantum 
optics itself.

\subsection{Non--Standard Quantum Model}

Similarly, from (\ref{eq:non-commutative Berry model}) we can define 
\begin{equation}
\label{eq:non-commutative modified Berry model}
H_{pJC}
=\tau_{+}\otimes a+\tau_{-}\otimes a^{\dagger} + 
\theta \tau_{3}\otimes {\bf 1}
=
\left(
  \begin{array}{cc}
    \theta      & a       \\
   -a^{\dagger} & -\theta
  \end{array}
\right),\quad [a,a^{\dagger}]={\bf 1}
\end{equation}
by replacing $\{\sigma_{j}\}$ with $\{\tau_{j}\}$. In this case this model is 
derived ``by hand". It satisfies the $su(1,1)$ like relation (see 
(\ref{eq:su(1,1)}))
\[
{\bf J}H_{pJC}{\bf J}=H_{pJC}^{\dagger}\quad ;\quad 
{\bf J}=
\left(
  \begin{array}{cc}
    {\bf 1} &          \\
            & -{\bf 1}
  \end{array}
\right)=J\otimes {\bf 1}.
\]

$H_{pJC}$ can be considered as a non-commutative version of $H_{pB}$ 
under the correspondence 
$
a\ \longleftrightarrow\ x-iy,\ a^{\dagger}\ \longleftrightarrow\ x+iy \ 
\mbox{and}\ \theta\ \longleftrightarrow\ z
$ : 
\begin{equation}
\label{eq:noncompact-type}
H_{pB}
=
\left(
  \begin{array}{cc}
    z     & x-iy \\
  -(x+iy) & -z
  \end{array}
\right),\ [x-iy,x+iy]=0 \ \longrightarrow \
H_{pJC}
=
\left(
  \begin{array}{cc}
    \theta      & a       \\
   -a^{\dagger} & -\theta
  \end{array}
\right),\ [a,a^{\dagger}]={\bf 1}.
\end{equation}

\vspace{3mm}
{\bf A comment is in order}.\quad In place of the Hamiltonian 
(\ref{eq:hamiltonian}) we can consider the following pseudo Hamiltonian 
\begin{equation}
\label{eq:pseudo-hamiltonian}
H_{p}=
\omega 1_{2}\otimes a^{\dagger}a + \frac{\Delta}{2} \tau_{3}\otimes {\bf 1} 
+ g\left(\tau_{+}\otimes a+\tau_{-}\otimes a^{\dagger} \right)
\end{equation}
by replacing $\{\sigma_{+},\sigma_{-},\sigma_{3}\}$ with 
$\{\tau_{+},\tau_{-},\tau_{3}\}$. This is not hermite (namely, not a 
convensional one), so we don't know whether this model is useful or not 
at the current time. It is interesting to note that the model has been 
considered by \cite{Mandal}.

In a forthcoming paper we will extend this ``Hamiltonian" and 
determine its structure in detail like \cite{papers}.

\section{Non--Commutative Hopf and Pseudo Hopf Bundles}

In this section we construct a non--commutative version of the Hopf and pseudo 
Hopf bundles by making (\ref{eq:non-commutative Berry model}) and 
(\ref{eq:non-commutative modified Berry model}) diagonal, which is 
a ``natural" extension in the section 3. 

First of all let us recall a Fock space. 
For $a$ and $a^\dagger$ we set $N\equiv a^{\dagger}a$ which is called 
the number operator, then we have 
\begin{equation}
  \label{eq:2-0}
  [N,a^\dagger]=a^{\dagger},\quad
  [N,a]=-a,\quad
  [a^\dagger, a]=-\mathbf{1}.
\end{equation}
Let ${\cal F}$ be the Fock space generated by $\{a,a^{\dagger},N\}$
\begin{equation}
  \label{eq:2-1}
{\cal F}=\mbox{Vect}_{\fukuso}\{\ket{0},\ket{1},\cdots,\ket{n},\cdots \}.
\end{equation}
The actions of $a$ and $a^{\dagger}$ on ${\cal F}$ are given by
\begin{equation}
  \label{eq:2-2}
  a\ket{n} = \sqrt{n}\ket{n-1},\quad
  a^{\dagger}\ket{n} = \sqrt{n+1}\ket{n+1},\quad
  N\ket{n} = n\ket{n}
\end{equation}
where $\ket{0}$ is a normalized vacuum ($a\ket{0}=0\  {\rm and}\  
\langle{0}\vert{0}\rangle = 1$). From (\ref{eq:2-2})
state $\ket{n}$ for $n \geq 1$ are given by
\begin{equation}
  \label{eq:2-3}
  \ket{n} = \frac{(a^{\dagger})^{n}}{\sqrt{n!}}\ket{0}\ .
\end{equation}
These states satisfy the orthogonality and completeness conditions 
\begin{equation}
  \label{eq:2-4}
   \langle{m}\vert{n}\rangle = \delta_{mn}\quad \mbox{and}\quad 
   \sum_{n=0}^{\infty}\ket{n}\bra{n} = \mathbf{1}. 
\end{equation}

\subsection{Non--Commutative Hopf Bundle}

First we make the Hamiltonian (\ref{eq:non-commutative Berry model}) 
diagonal like in Section 2 and research whether ``Dirac strings" exist or not 
in this non--commutative model, which is very interesting from not only 
quantum optical but also mathematical point of view.

It is easy to see 
\begin{equation}
\label{eq:non-commutative decomposition}
H_{JC}
=
\left(
  \begin{array}{cc}
    \theta      & a       \\
    a^{\dagger} & -\theta
  \end{array}
\right)
=
\left(
  \begin{array}{cc}
    1 &                                 \\
      & a^{\dagger}\frac{1}{\sqrt{N+1}}
  \end{array}
\right)
\left(
  \begin{array}{cc}
     \theta   & \sqrt{N+1} \\
   \sqrt{N+1} & -\theta
  \end{array}
\right)
\left(
  \begin{array}{cc}
    1 &                       \\
      & \frac{1}{\sqrt{N+1}}a
  \end{array}
\right)
\end{equation}
from \cite{FHKSW}. 
Then the middle matrix in the right hand side can be considered as a classical 
one, so we can diagonalize it easily 
\begin{equation}
\label{eq:middle matrix decomposition}
\left(
  \begin{array}{cc}
     \theta   & \sqrt{N+1} \\
   \sqrt{N+1} & -\theta
  \end{array}
\right)
=
\left\{
\begin{array}{l}
 A_{I}
 \left(
  \begin{array}{cc}
    R(N+1) &         \\
           & -R(N+1)
  \end{array}
 \right)
 A_{I}^{\dagger}\\
 A_{II}
 \left(
  \begin{array}{cc}
    R(N+1) &         \\
           & -R(N+1)
  \end{array}
 \right)
 A_{II}^{\dagger} 
\end{array}
\right.
\end{equation}
where 
\[
R(N)=\sqrt{N+\theta^{2}} 
\]
and $A_{I}$, $A_{II}$ are defined by 
\begin{eqnarray}
\label{eq:classical unitary-i}
A_{I}&=&
\frac{1}{\sqrt{2R(N+1)(R(N+1)+\theta)}}
\left(
  \begin{array}{cc}
   R(N+1)+\theta & -\sqrt{N+1}   \\
   \sqrt{N+1}    & R(N+1)+\theta
  \end{array}
\right),  \\
\label{eq:classical unitary-ii}
A_{II}&=&
\frac{1}{\sqrt{2R(N+1)(R(N+1)-\theta)}}
\left(
  \begin{array}{cc}
      \sqrt{N+1}   & -R(N+1)+\theta \\
     R(N+1)-\theta & \sqrt{N+1} 
  \end{array}
\right). 
\end{eqnarray}
Now let us rewrite (\ref{eq:non-commutative decomposition}) by making use of 
(\ref{eq:middle matrix decomposition}) with (\ref{eq:classical unitary-i}). 
Inserting the identity 
\[
\left(
  \begin{array}{cc}
    1 &                       \\
      & \frac{1}{\sqrt{N+1}}a
  \end{array}
\right)
\left(
  \begin{array}{cc}
    1 &                                 \\
      & a^{\dagger}\frac{1}{\sqrt{N+1}}
  \end{array}
\right)
=
\left(
  \begin{array}{cc}
    1 &   \\
      & 1
  \end{array}
\right)
\]
gives 
\begin{eqnarray}
H_{JC}=
&&\left(
  \begin{array}{cc}
    1 &                                 \\
      & a^{\dagger}\frac{1}{\sqrt{N+1}}
  \end{array}
\right)
A_{I}
\left(
  \begin{array}{cc}
    R(N+1) &         \\
           & -R(N+1)
  \end{array}
\right)
A_{I}^{\dagger}
\left(
  \begin{array}{cc}
    1 &                       \\
      & \frac{1}{\sqrt{N+1}}a
  \end{array}
\right)    \nonumber    \\
=
&&
\left(
  \begin{array}{cc}
    1 &                                 \\
      & a^{\dagger}\frac{1}{\sqrt{N+1}}
  \end{array}
\right)
A_{I}
\left(
  \begin{array}{cc}
    1 &                       \\
      & \frac{1}{\sqrt{N+1}}a
  \end{array}
\right)
\left(
  \begin{array}{cc}
    1 &                                 \\
      & a^{\dagger}\frac{1}{\sqrt{N+1}}
  \end{array}
\right)
\left(
  \begin{array}{cc}
    R(N+1) &         \\
           & -R(N+1)
  \end{array}
\right)\times   \nonumber  \\
&&
\left(
  \begin{array}{cc}
    1 &                       \\
      & \frac{1}{\sqrt{N+1}}a
  \end{array}
\right)
\left(
  \begin{array}{cc}
    1 &                                 \\
      & a^{\dagger}\frac{1}{\sqrt{N+1}}
  \end{array}
\right)
A_{I}^{\dagger}
\left(
  \begin{array}{cc}
    1 &                       \\
      & \frac{1}{\sqrt{N+1}}a
  \end{array}
\right)         \nonumber  \\
=
&&
U_{I}
\left(
  \begin{array}{cc}
    R(N+1) &         \\
           & -R(N)
  \end{array}
\right)
U_{I}^{\dagger}, 
\end{eqnarray}
where
\begin{eqnarray}
\label{eq:V-I}
U_{I}
&=&
\left(
  \begin{array}{cc}
    \frac{1}{\sqrt{2R(N+1)(R(N+1)+\theta)}} &      \\
           & \frac{1}{\sqrt{2R(N)(R(N)+\theta)}}
  \end{array}
\right)
\left(
  \begin{array}{cc}
      R(N+1)+\theta & -a            \\
      a^{\dagger}   &  R(N)+\theta
  \end{array}
\right)         \nonumber \\
&=&
\left(
  \begin{array}{cc}
      R(N+1)+\theta & -a            \\
      a^{\dagger}   &  R(N)+\theta
  \end{array}
\right)  
\left(
  \begin{array}{cc}
    \frac{1}{\sqrt{2R(N+1)(R(N+1)+\theta)}} &      \\
           & \frac{1}{\sqrt{2R(N)(R(N)+\theta)}}
  \end{array}
\right).
\end{eqnarray}

Similarly, we can rewrite (\ref{eq:non-commutative decomposition}) by making 
use of (\ref{eq:middle matrix decomposition}) with 
(\ref{eq:classical unitary-ii}). 
By inserting the identity 
\[
\left(
  \begin{array}{cc}
    \frac{1}{\sqrt{N+1}}a &    \\
                          & 1
  \end{array}
\right)
\left(
  \begin{array}{cc}
    a^{\dagger}\frac{1}{\sqrt{N+1}} &    \\
                                    & 1
  \end{array}
\right)
=
\left(
  \begin{array}{cc}
    1 &   \\
      & 1
  \end{array}
\right)
\]
we obtain 
\begin{equation}
H_{JC}=
U_{II}
\left(
  \begin{array}{cc}
    R(N) &          \\
         & -R(N+1)
  \end{array}
\right)
U_{II}^{\dagger}, 
\end{equation}
where 
\begin{eqnarray}
\label{eq:V-II}
U_{II}
&=&
\left(
  \begin{array}{cc}
    \frac{1}{\sqrt{2R(N+1)(R(N+1)-\theta)}} &     \\
           & \frac{1}{\sqrt{2R(N)(R(N)-\theta)}}
  \end{array}
\right)
\left(
  \begin{array}{cc}
          a       & -R(N+1)+\theta  \\
      R(N)-\theta &  a^{\dagger} 
  \end{array}
\right)        \nonumber \\
&=&
\left(
  \begin{array}{cc}
          a       & -R(N+1)+\theta  \\
      R(N)-\theta &  a^{\dagger} 
  \end{array}
\right) 
\left(
  \begin{array}{cc}
    \frac{1}{\sqrt{2R(N)(R(N)-\theta)}} &             \\
           & \frac{1}{\sqrt{2R(N+1)(R(N+1)-\theta)}}
  \end{array}
\right).
\end{eqnarray}

Tidying up these we have 
\begin{equation}
\label{eq:final-decomposition}
H_{JC}=
\left\{
\begin{array}{l}
U_{I}
\left(
  \begin{array}{cc}
    R(N+1) &         \\
           & -R(N)
  \end{array}
\right)
U_{I}^{\dagger}\\
U_{II}
\left(
  \begin{array}{cc}
    R(N) &         \\
         & -R(N+1)
  \end{array}
\right)
U_{II}^{\dagger}
\end{array}
\right.
\end{equation}
with $U_{I}$ and $U_{II}$ above. From the equations 
\[
R(N+1)\ket{0}=\sqrt{1+\theta^{2}}>\theta,\quad
R(N)\ket{0}=\sqrt{\theta^{2}}=|\theta|
\]
we know 
\[
\left(R(N)\pm \theta\right)\ket{0}=\left(|\theta|\pm \theta\right)\ket{0},
\]
so the strings corresponding to Dirac ones exist in only states 
${\cal F}\times \{\ket{0}\} \cup \{\ket{0}\}\times {\cal F}$ where ${\cal F}$ 
is the Fock space, while in other excited states 
${\cal F}\times {\cal F}\setminus 
{\cal F}\times \{\ket{0}\} \cup \{\ket{0}\}\times {\cal F}$ they don't exist
\footnote{We have identified ${\cal F}\times {\cal F}$ with the space of 
$2$--component vectors over ${\cal F}$}, see the figure 3. 
The phenomenon is very interesting. 
For simplicity we again call these strings Dirac ones in the following. 

The ``parameter space" of $H_{JC}$ can be identified with 
${\cal F}\times {\cal F}\times \real \ni (*, *, \theta)$, so the domains 
$D_{I}$ of $U_{I}$ and $D_{II}$ of $U_{II}$ are respectively 
\begin{eqnarray}
D_{I}&=&{\cal F}\times {\cal F}\times \real - 
{\cal F}\times \{\ket{0}\}\times \real_{\leq 0},  \\
D_{II}&=&{\cal F}\times {\cal F}\times \real - 
\left({\cal F}\times \{\ket{0}\} \cup \{\ket{0}\}\times {\cal F}\right)
\times \real_{\geq 0}
\end{eqnarray}
by (\ref{eq:V-I}) and (\ref{eq:V-II}). We note that 
\[
D_{I}\cup D_{II}={\cal F}\times {\cal F}\times \real - 
{\cal F}\times \{\ket{0}\}\times \{\theta=0\}.
\]
%

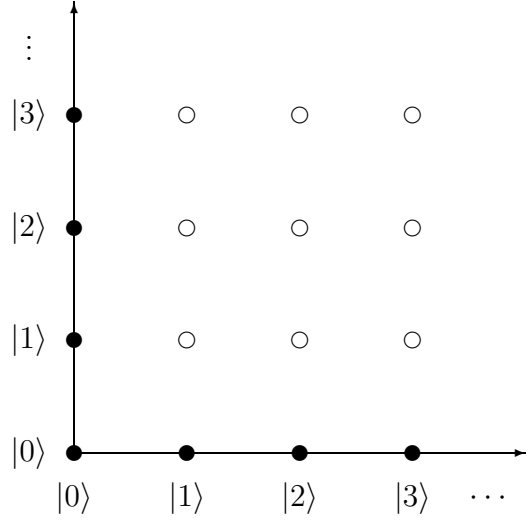
\begin{figure}
\begin{center}
\setlength{\unitlength}{1mm} 
\begin{picture}(80,70)
\put(10,10){\vector(1,0){60}}
\put(10,10){\vector(0,1){60}}
\put(10,10){\circle*{2}}
\put(25,10){\circle*{2}}
\put(40,10){\circle*{2}}
\put(55,10){\circle*{2}}
%
\put(10,25){\circle*{2}}
\put(10,40){\circle*{2}}
\put(10,55){\circle*{2}}
\put(25,25){\circle{2}}
\put(25,40){\circle{2}}
\put(25,55){\circle{2}}
\put(40,25){\circle{2}}
\put(40,40){\circle{2}}
\put(40,55){\circle{2}}
\put(55,25){\circle{2}}
\put(55,40){\circle{2}}
\put(55,55){\circle{2}}
\put(7,1){\makebox(6,6)[c]{$|0\rangle$}}
\put(22,1){\makebox(6,6)[c]{$|1\rangle$}}
\put(37,1){\makebox(6,6)[c]{$|2\rangle$}}
\put(52,1){\makebox(6,6)[c]{$|3\rangle$}}
\put(62,1){\makebox(6,6)[c]{$\cdots$}}
\put(1,7){\makebox(6,6)[c]{$|0\rangle$}}
\put(1,22){\makebox(6,6)[c]{$|1\rangle$}}
\put(1,37){\makebox(6,6)[c]{$|2\rangle$}}
\put(1,52){\makebox(6,6)[c]{$|3\rangle$}}
\put(1,62){\makebox(6,6)[c]{$\vdots$}}
\end{picture}
\vspace{-5mm}
\caption{The bases of ${\cal F}\times {\cal F}$. The black circle means 
bases giving Dirac strings, while the white one don't.}
\end{center}
\end{figure}

Then the transition ``function" (operator) is given by 
\[
\Phi_{JC}=
\left(
  \begin{array}{cc}
    a\frac{1}{\sqrt{N}} &                               \\
                        & \frac{1}{\sqrt{N}}a^{\dagger}
  \end{array}
\right)
=
\left(
  \begin{array}{cc}
    \frac{1}{\sqrt{N+1}}a &                                 \\
                          & a^{\dagger}\frac{1}{\sqrt{N+1}}
  \end{array}
\right).
\]
Therefore the system 
\begin{equation}
\label{eq:noncommutative-Hopf-bundle}
\left\{(U_{I},D_{I}), (U_{II},D_{II}), \Phi_{JC}, D_{I}\cup D_{II}\right\}
\end{equation}
is a non-commutative version of the Hopf bundle (\ref{eq:Hopf-bundle}). 
The projector in this case becomes
\begin{eqnarray}
\label{eq:quantum-projector}
P_{JC}&=&
U_{I}
\left(
  \begin{array}{cc}
    {\bf 1} &         \\
            & {\bf 0}
  \end{array}
\right)
U_{I}^{\dagger}
=
U_{II}
\left(
  \begin{array}{cc}
    {\bf 1} &         \\
            & {\bf 0}
  \end{array}
\right)
U_{II}^{\dagger}  \nonumber \\
&=&
\left\{
\begin{array}{l}
\left(
  \begin{array}{cc}
    \frac{1}{2R(N+1)} &                  \\
                      & \frac{1}{2R(N)} 
  \end{array}
\right)
\left(
  \begin{array}{cc}
      R(N+1)+\theta &  a            \\
      a^{\dagger}   &  R(N)-\theta
  \end{array}
\right)      \\
\left(
  \begin{array}{cc}
      R(N+1)+\theta &  a            \\
      a^{\dagger}   &  R(N)-\theta
  \end{array}
\right)
\left(
  \begin{array}{cc}
    \frac{1}{2R(N+1)} &                  \\
                      & \frac{1}{2R(N)} 
  \end{array}
\right).
\end{array}
\right.
\end{eqnarray}
Note that the projector $P_{JC}$ is not defined on 
${\cal F}\times \{\ket{0}\}\times \{\theta=0\}$ = ${\cal F}\times
{\cal F}\times \real - D_{I}\cup D_{II}$. 

\par \vspace{5mm}
A comment is in order. From (\ref{eq:quantum-projector}) we obtain a quantum 
version of (classical) spectral decomposition 
(a ``quantum spectral decomposition" by Suzuki \cite{TS}) 
\begin{equation}
\label{eq:quantum-spectral-decomposition}
H_{JC}=
\left(
  \begin{array}{cc}
    R(N+1) &       \\
           & R(N)
  \end{array}
\right)
P_{JC}
-
\left(
  \begin{array}{cc}
    R(N+1) &       \\
           & R(N)
  \end{array}
\right)
({\bf 1}_{2}-P_{JC}).
\end{equation}

As a bonus of the decomposition let us rederive the calculation of 
$\mbox{e}^{-igtH_{JC}}$ which has been given in \cite{MS}. 
The result is 
\begin{equation}
\mbox{e}^{-igtH_{JC}}=
\left(
  \begin{array}{cc}
   \mbox{cos}(tgR(N+1))-i\theta\frac{\mbox{sin}(tgR(N+1))}{R(N+1)}& 
   -i\frac{\mbox{sin}(tgR(N+1))}{R(N+1)}a       \\
   -i\frac{\mbox{sin}(tgR(N))}{R(N)}a^{\dagger}
   & \mbox{cos}(tgR(N))+i\theta\frac{\mbox{sin}(tgR(N))}{R(N)}
  \end{array}
\right)
\end{equation}
by making use of (\ref{eq:final-decomposition}) (or 
(\ref{eq:quantum-spectral-decomposition})). 
We leave it to the readers.

\subsection{Non--Commutative Pseudo Hopf Bundle}

Similarly, we make the Hamiltonian 
(\ref{eq:non-commutative modified Berry model}) diagonal like in the preceding 
subsection to study what a non--commutative version of Dirac strings is.

It is easy to see 
\begin{equation}
\label{eq:non-commutative decomposition noncompact}
H_{pJC}
=
\left(
  \begin{array}{cc}
    \theta      & a       \\
   -a^{\dagger} & -\theta
  \end{array}
\right)
=
\left(
  \begin{array}{cc}
    1 &                                 \\
      & a^{\dagger}\frac{1}{\sqrt{N+1}}
  \end{array}
\right)
\left(
  \begin{array}{cc}
     \theta   & \sqrt{N+1} \\
  -\sqrt{N+1} & -\theta
  \end{array}
\right)
\left(
  \begin{array}{cc}
    1 &                       \\
      & \frac{1}{\sqrt{N+1}}a
  \end{array}
\right).
\end{equation}
The middle matrix in the right hand side can be considered as a classical 
one, so we can diagonalize it easily 
\begin{equation}
\label{eq:middle matrix decomposition non-compact}
\left(
  \begin{array}{cc}
     \theta   & \sqrt{N+1} \\
  -\sqrt{N+1} & -\theta
  \end{array}
\right)
=
 B
 \left(
  \begin{array}{cc}
    S(N+1) &         \\
           & -S(N+1)
  \end{array}
 \right)
 B^{-1}
\end{equation}
where 
\[
S(N)=\sqrt{\theta^{2}-N} 
\]
and $B$ defined by 
\begin{equation}
\label{eq:classical unitary noncompact}
B=
\frac{1}{\sqrt{2S(N+1)(\theta+S(N+1))}}
\left(
  \begin{array}{cc}
   S(N+1)+\theta & -\sqrt{N+1}   \\
  -\sqrt{N+1}    & S(N+1)+\theta
  \end{array}
\right).  
\end{equation}

In this case the situation changes in a drastic manner. Since 
$S(N+1)=\sqrt{\theta^{2}-(N+1)}$ where $N=a^{\dagger}a$ is the number operator, 
it is clear that only a restricted subspace of the Fock space ${\cal{F}}$ 
\[
{\cal{F}}_{n}=\mbox{Vect}_{\fukuso}\{\ket{0},\ket{1},\cdots,\ket{n-1}\}
\]
is available if $n <\theta^{2} \leq n+1$. Moreover, if 
$0 < \theta^{2} \leq 1$ there is no subspace that $S(N+1)$ is defined !

Similarly in the preceding subsection we have 
\begin{equation}
\label{eq:final-decomposition noncompact}
H_{pJC}=
V
\left(
  \begin{array}{cc}
    S(N+1) &         \\
           & -S(N)
  \end{array}
\right)
V^{-1}\\
\end{equation}
with $V$ defined by 
\begin{eqnarray}
\label{eq:W-I}
V
&=&
\left(
  \begin{array}{cc}
    \frac{1}{\sqrt{2S(N+1)(S(N+1)+\theta)}} &      \\
           & \frac{1}{\sqrt{2S(N)(S(N)+\theta)}}
  \end{array}
\right)
\left(
  \begin{array}{cc}
      S(N+1)+\theta & -a            \\
     -a^{\dagger}   &  S(N)+\theta
  \end{array}
\right)         \nonumber \\
&=&
\left(
  \begin{array}{cc}
      R(N+1)+\theta & -a            \\
     -a^{\dagger}   &  R(N)+\theta
  \end{array}
\right)  
\left(
  \begin{array}{cc}
    \frac{1}{\sqrt{2S(N+1)(S(N+1)+\theta)}} &      \\
           & \frac{1}{\sqrt{2S(N)(S(N)+\theta)}}
  \end{array}
\right).
\end{eqnarray}
We note that $V$ above satisfies the relation 
\[
V^{\dagger}{\bf J}V={\bf J},\quad  \mbox{where}\quad
{\bf J}=
\left(
  \begin{array}{cc}
    {\bf 1} &          \\
            & -{\bf 1}
  \end{array}
\right)
\Longrightarrow
V^{-1}={\bf J}V^{\dagger}{\bf J}.
\]
The ``parameter space" of $H_{pJC}$ can be identified with 
\[
\bigcup_{n\in \futon}
{\cal F}_{n}\times {\cal F}_{n+1}\times 
\{\theta \in \real_{>0}\ |\ n < \theta^{2} \leq n+1\} 
\ni (*, *, \theta).
\]

The projector in this case becomes
\begin{eqnarray}
\label{eq:quantum-projector noncompact}
Q_{pJC}&=&
V
\left(
  \begin{array}{cc}
    {\bf 1} &         \\
            & {\bf 0}
  \end{array}
\right)
V^{-1}  \nonumber \\
&=&
\left\{
\begin{array}{l}
\left(
  \begin{array}{cc}
    \frac{1}{2S(N+1)} &                  \\
                      & \frac{1}{2S(N)} 
  \end{array}
\right)
\left(
  \begin{array}{cc}
      S(N+1)+\theta &  a            \\
     -a^{\dagger}   &  S(N)-\theta
  \end{array}
\right)      \\
\left(
  \begin{array}{cc}
      S(N+1)+\theta &  a            \\
     -a^{\dagger}   &  S(N)-\theta
  \end{array}
\right)
\left(
  \begin{array}{cc}
    \frac{1}{2S(N+1)} &                  \\
                      & \frac{1}{2S(N)} 
  \end{array}
\right).
\end{array}
\right.
\end{eqnarray}
It is easy to see the relations 
\[
Q_{pJC}^{2}=Q_{pJC},\quad {\bf J}Q_{pJC}{\bf J}=Q_{pJC}^{\dagger}.
\]

\par \vspace{5mm}
A comment is in order. From (\ref{eq:quantum-projector noncompact}) we obtain 
a quantum version of (classical) spectral decomposition
\begin{equation}
\label{eq:quantum-spectral-decomposition noncompact}
H_{pJC}=
\left(
  \begin{array}{cc}
    S(N+1) &       \\
           & S(N)
  \end{array}
\right)
Q_{pJC}
-
\left(
  \begin{array}{cc}
    S(N+1) &       \\
           & S(N)
  \end{array}
\right)
({\bf 1}_{2}-Q_{pJC}).
\end{equation}

As a bonus of the decomposition let us rederive the calculation of 
$\mbox{e}^{-igtH_{pJC}}$ which seems to be new. 
The result is 
\begin{equation}
\mbox{e}^{-igtH_{pJC}}=
\left(
  \begin{array}{cc}
   \mbox{cos}(tgS(N+1))-i\theta\frac{\mbox{sin}(tgS(N+1))}{S(N+1)}& 
   -i\frac{\mbox{sin}(tgS(N+1))}{S(N+1)}a       \\
    i\frac{\mbox{sin}(tgS(N))}{S(N)}a^{\dagger}
   & \mbox{cos}(tgS(N))+i\theta\frac{\mbox{sin}(tgS(N))}{S(N)}
  \end{array}
\right)
\end{equation}
by making use of (\ref{eq:final-decomposition noncompact}) (or 
(\ref{eq:quantum-spectral-decomposition noncompact})). 
We leave it to the readers. 

We note once more that $\mbox{e}^{-igtH_{pJC}}$ is not unitary, but satisfies 
the relation 
\[
\left(\mbox{e}^{-igtH_{pJC}}\right)^{\dagger}{\bf J}\mbox{e}^{-igtH_{pJC}}
={\bf J}.
\]

\section{Non--Commutative Veronese and Pseudo Veronese \\ Mappings}

In this section we construct a non--commutative version of the 
(classical) Veronese Mapping and its noncompact counterpart.

\subsection{Non--Commutative Veronese Mapping}

Let us make a brief review of the Veronese mapping. The map 
\[
{\fukuso}P^{1} \longrightarrow {\fukuso}P^{n}
\]
is defined as 
\[
[z_{1}:z_{2}] \longrightarrow 
\left[z_{1}^{n}:\sqrt{{}_nC_1}z_{1}^{n-1}z_{2}:\cdots:
\sqrt{{}_nC_j}z_{1}^{n-j}z_{2}^{j}:\cdots:
\sqrt{{}_nC_{n-1}}z_{1}z_{2}^{n-1}:z_{2}^{n}\right] 
\]
by making use of the homogeneous coordinate, see Appendix A.
We also have another expression of this map by using 
\[
S_{\fukuso}^{1} \longrightarrow S_{\fukuso}^{n}\ :\ 
v_{1}\equiv
\left(
  \begin{array}{c}
    z_{1} \\
    z_{2}
  \end{array}
\right)
\ \longrightarrow \
v_{n}\equiv
\left(
  \begin{array}{c}
    z_{1}^{n} \\
    \sqrt{{}_nC_1}z_{1}^{n-1}z_{2} \\
    \vdots \\
    \sqrt{{}_nC_j}z_{1}^{n-j}z_{2}^{j} \\
    \vdots \\
    \sqrt{{}_nC_{n-1}}z_{1}z_{2}^{n-1} \\
    z_{2}^{n}
  \end{array}
\right),
\quad |z_{1}|^{2}+|z_{2}|^{2}=1
\]
where $S_{\fukuso}^{m}=\left\{(w_{1},w_{2},\cdots,w_{m+1})^{\mbox{T}} \in 
\fukuso^{m+1}\ |\ \sum_{j=1}^{m+1}|w_{j}|^{2}=1\right\} \cong S^{2m+1}$ and 
${\fukuso}P^{m}=S_{\fukuso}^{m}/U(1)$. 
Then the Veronese mapping is also written as
\[
{\fukuso}P^{1} \longrightarrow {\fukuso}P^{n}\ :\ 
P_{1}=v_{1}v_{1}^{\dagger}
\ \longmapsto \
P_{n}=v_{n}v_{n}^{\dagger}.
\]
by using projectors, which is easy to understand.
\par \noindent
Moreover, the local map ($z\equiv z_{2}/z_{1}$) is given as 
\[
\fukuso \longrightarrow \fukuso^{n} : 
z \longrightarrow \
\left(
  \begin{array}{c}
    \sqrt{{}_nC_1}z \\
    \vdots \\
    \sqrt{{}_nC_j}z^{j} \\
    \vdots \\
    \sqrt{{}_nC_{n-1}}z^{n-1} \\
    z^{n}
  \end{array}
\right).
\]
See the following picture as a whole.

\vspace{-5mm}
\setlength{\unitlength}{1mm} 
\begin{center}
\begin{picture}(70,60)
\put(20,26){\vector(1,0){30}}
\put(20,50){\vector(1,0){30}}
\put(20, 2){\vector(1,0){30}}
\put(12,45){\vector(0,-1){15}}
\put(57,45){\vector(0,-1){15}}
\put(10,23){\makebox(6,6)[c]{${\fukuso}P^{1}$}}
\put(55,23){\makebox(6,6)[c]{${\fukuso}P^{n}$}}
\put(10,47){\makebox(6,6)[c]{$S_{\fukuso}^{1}$}}
\put(55,47){\makebox(6,6)[c]{$S_{\fukuso}^{n}$}}
\put(10,-1){\makebox(6,6)[c]{${\fukuso}$}}
\put(55,-1){\makebox(6,6)[c]{${\fukuso}^{n}$}}
\put(12, 6){\vector(0, 1){15}}
\put(57, 6){\vector(0, 1){15}}
\end{picture}
\end{center}

Next we want to consider a non--commutative version of the map. 
If we set 
\begin{equation}
{\cal A}\equiv 
\left(
  \begin{array}{c}
   X_{0} \\
   Y_{0}
  \end{array}
\right)
=
\left(
  \begin{array}{c}
    \frac{R(N+1)+\theta}{\sqrt{2R(N+1)(R(N+1)+\theta)}} \\
    \frac{1}{\sqrt{2R(N)(R(N)+\theta)}}a^{\dagger}
  \end{array}
\right)
\end{equation}
from $U_{I}$ in (\ref{eq:V-I}), then 
\[
{\cal A}^{\dagger}{\cal A}=X_{0}^{2}+Y_{0}^{\dagger}Y_{0}={\bf 1}
\quad \mbox{and}\quad 
Y_{0}X_{0}^{-1}=\frac{1}{R(N)+\theta}a^{\dagger}\equiv Z.
\]
That is, ${\cal A}=(X_{0},Y_{0})^{T}$ is a non--commutative sphere and 
$Z$ is a kind of ``stereographic projection" of the sphere. 
It is easy to see the following 
\begin{equation}
\label{eq:easy-formula}
{\bf 1}+Z^{\dagger}Z=\frac{2R(N+1)}{R(N+1)+\theta}=X_{0}^{-2}
\Longrightarrow 
X_{0}=\left({\bf 1}+Z^{\dagger}Z\right)^{-1/2}.
\end{equation}

Here let us introduce new notations for the following. For $j\geq 0$ we set 
\begin{eqnarray}
\label{eq:def-X}
X_{-j}&=&\frac{R(N+1-j)+\theta}{\sqrt{2R(N+1-j)(R(N+1-j)+\theta)}}, \\
\label{eq:def-Y}
Y_{-j}&=&\sqrt{\frac{N-j}{N}}\frac{1}{\sqrt{2R(N-j)(R(N-j)+\theta)}}a^{\dagger}
.
\end{eqnarray}
We list some useful formulas.
\begin{equation}
\label{eq:useful-formulas}
X_{-j}^{2}+Y_{-j}^{\dagger}Y_{-j}={\bf 1}
\quad \mbox{and}\quad
Y_{-j}^{\dagger}Y_{-j}=Y_{-j+1}Y_{-j+1}^{\dagger}\quad 
\mbox{for}\quad j\geq 0.
\end{equation}

Now we are in a position to define a quantum version of the Veronese mapping 
which plays a very important role in ``classical" Mathematics.

\begin{equation}
\label{eq:Quantum-Mapping}
{\cal A}=
\left(
  \begin{array}{c}
   X_{0} \\
   Y_{0}
  \end{array}
\right)
\longrightarrow 
{\cal A}_{n}=
\left(
  \begin{array}{c}
   X_{0}^{n} \\
   \sqrt{{}_nC_1}Y_{0}X_{0}^{n-1} \\
   \vdots \\
   \sqrt{{}_nC_j}Y_{-(j-1)}Y_{-(j-2)}\cdots Y_{-1}Y_{0}X_{0}^{n-j} \\
   \vdots \\
   \sqrt{{}_nC_{n-1}}Y_{-(n-2)}Y_{-(n-3)}\cdots Y_{-1}Y_{0}X_{0} \\
   Y_{-(n-1)}Y_{-(n-2)}Y_{-(n-3)}\cdots Y_{-1}Y_{0}
  \end{array}
\right).
\end{equation}
Then it is not difficult to see 
\[
{\cal A}_{n}^{\dagger}{\cal A}_{n}=
\left(X_{0}^{2}+Y_{0}^{\dagger}Y_{0}\right)^{n}
=\left({\cal A}^{\dagger}{\cal A}\right)^{n}={\bf 1}.
\]
From this we can define the projectors which correspond to projective spaces 
like 
\begin{equation}
{\cal P}_{n}={\cal A}_{n}{\cal A}_{n}^{\dagger},\quad 
{\cal P}_{1}={\cal A}{\cal A}^{\dagger},
\end{equation}
so the map 
\begin{equation}
\label{eq:Quantum-Veronese-Mapping}
{\cal P}_{1}\ \longrightarrow \ {\cal P}_{n}
\end{equation}
is a non-commutative version of the Veronese mapping.

Next, we define a local ``coordinate" of the Veronese mapping defined above. 
\begin{eqnarray*}
{\cal A}_{n}
&=&
\left(
  \begin{array}{c}
   {\bf 1} \\
   \sqrt{{}_nC_1}Y_{0}X_{0}^{-1} \\
   \vdots \\
   \sqrt{{}_nC_j}Y_{-(j-1)}Y_{-(j-2)}\cdots Y_{-1}Y_{0}X_{0}^{-j} \\
   \vdots \\
   \sqrt{{}_nC_{n-1}}Y_{-(n-2)}Y_{-(n-3)}\cdots Y_{-1}Y_{0}X_{0}^{-(n-1)} \\
   Y_{-(n-1)}Y_{-(n-2)}Y_{-(n-3)}\cdots Y_{-1}Y_{0}X_{0}^{-n}
  \end{array}
\right)
X_{0}^{n} \\
&&{} \\
&=&\cdots \\
&&{} \\
&=&
\left(
  \begin{array}{c}
   {\bf 1} \\
   \sqrt{{}_nC_1}Y_{0}X_{0}^{-1} \\
   \vdots \\
   \sqrt{{}_nC_j}Y_{-(j-1)}X_{-(j-1)}^{-1}Y_{-(j-2)}X_{-(j-2)}^{-1}
   \cdots Y_{-1}X_{-1}^{-1}Y_{0}X_{0}^{-1} \\
   \vdots \\
   \sqrt{{}_nC_{n-1}}Y_{-(n-2)}X_{-(n-2)}^{-1}Y_{-(n-3)}X_{-(n-3)}^{-1}
   \cdots Y_{-1}X_{-1}^{-1}Y_{0}X_{0}^{-1} \\
   Y_{-(n-1)}X_{-(n-1)}^{-1}Y_{-(n-2)}X_{-(n-2)}^{-1}Y_{-(n-3)}X_{-(n-3)}^{-1}
   \cdots Y_{-1}X_{-1}^{-1}Y_{0}X_{0}^{-1}
  \end{array}
\right)
X_{0}^{n}
\end{eqnarray*}
where we have used the relation 
\[
Y_{-j}X_{-k}^{-1}=X_{-(k+1)}^{-1}Y_{-j}
\]
due to $a^{\dagger}$ in $Y_{-j}$. Moreover, by (\ref{eq:def-X}) and 
(\ref{eq:def-Y})
\[
Y_{-j}X_{-j}^{-1}=
\sqrt{\frac{N-j}{N}}\frac{1}{R(N-j)+\theta}a^{\dagger}\equiv Z_{-j}
\quad \mbox{for}\quad j\geq 0.
\]
Note that $Z_{0}=Z$. Therefore by using (\ref{eq:easy-formula}) we have 
\begin{equation}
{\cal A}_{n}=
\left(
  \begin{array}{c}
   {\bf 1} \\
   \sqrt{{}_nC_1}Z_{0} \\
   \vdots \\
   \sqrt{{}_nC_j}Z_{-(j-1)}Z_{-(j-2)}\cdots Z_{-1}Z_{0} \\
   \vdots \\
   \sqrt{{}_nC_{n-1}}Z_{-(n-2)}Z_{-(n-3)}\cdots Z_{-1}Z_{0} \\
   Z_{-(n-1)}Z_{-(n-2)}Z_{-(n-3)}\cdots Z_{-1}Z_{0}
  \end{array}
\right)
\left({\bf 1}+Z_{0}^{\dagger}Z_{0}\right)^{-n/2}.
\end{equation}

\par \noindent
Now if we define 
\begin{equation}
{\cal Z}_{n}=
\left(
  \begin{array}{c}
   \sqrt{{}_nC_1}Z_{0} \\
   \vdots \\
   \sqrt{{}_nC_j}Z_{-(j-1)}Z_{-(j-2)}\cdots Z_{-1}Z_{0} \\
   \vdots \\
   \sqrt{{}_nC_{n-1}}Z_{-(n-2)}Z_{-(n-3)}\cdots Z_{-1}Z_{0} \\
   Z_{-(n-1)}Z_{-(n-2)}Z_{-(n-3)}\cdots Z_{-1}Z_{0}
  \end{array}
\right),
\end{equation}
then 
\[
{\cal A}_{n}=
\left(
  \begin{array}{c}
   {\bf 1} \\
   {\cal Z}_{n}
  \end{array}
\right)
\left({\bf 1}+Z_{0}^{\dagger}Z_{0}\right)^{-n/2}.
\]
and it is easy to show 
\[
{\bf 1}+{\cal Z}_{n}^{\dagger}{\cal Z}_{n}=
\left({\bf 1}+Z_{0}^{\dagger}Z_{0}\right)^{n},
\]
so we obtain 
\begin{eqnarray}
\label{eq:full-expression}
{\cal P}_{n}&=&{\cal A}_{n}{\cal A}_{n}^{\dagger}  \nonumber \\
&=&
\left(
  \begin{array}{c}
    {\bf 1} \\
    {\cal Z}_{n}
  \end{array}
\right)
\left({\bf 1}+{\cal Z}_{n}^{\dagger}{\cal Z}_{n}\right)^{-1}
\left({\bf 1},\ {\cal Z}_{n}^{\dagger}\right)       \nonumber \\
&=&
\left(
  \begin{array}{cc}
   \left({\bf 1}+{\cal Z}_{n}^{\dagger}{\cal Z}_{n}\right)^{-1} & 
   \left({\bf 1}+{\cal Z}_{n}^{\dagger}{\cal Z}_{n}\right)^{-1}
   {\cal Z}_{n}^{\dagger}    \\
   {\cal Z}_{n}\left({\bf 1}+{\cal Z}_{n}^{\dagger}{\cal Z}_{n}\right)^{-1} &
   {\cal Z}_{n}\left({\bf 1}+{\cal Z}_{n}^{\dagger}{\cal Z}_{n}\right)^{-1}
   {\cal Z}_{n}^{\dagger}
  \end{array}
\right)         \nonumber \\
&=&
\left(
  \begin{array}{cc}
    {\bf 1}      & -{\cal Z}_{n}^{\dagger} \\
    {\cal Z}_{n} & {\bf 1}
  \end{array}
\right)
\left(
  \begin{array}{cc}
    {\bf 1} &          \\
            & {\bf 0}
  \end{array}
\right)
\left(
  \begin{array}{cc}
    {\bf 1}       & -{\cal Z}_{n}^{\dagger} \\
     {\cal Z}_{n} & {\bf 1}
  \end{array}
\right)^{-1}.
\end{eqnarray}
This is the Oike expression in \cite{Fujii}, see also Appendix B. 

\par \vspace{3mm}
{\bf A comment is in order}.\ Two of important properties which the classical 
Veronese mapping has are 
\begin{enumerate}
\item The Veronese mapping $\fukuso P^{1} \longrightarrow \fukuso P^{n}$ 
has the { \bf mapping degree} $n$
\item The Veronese surface (which is the image of Veronese mapping) is 
a {\bf minimal surface} in $\fukuso P^{n}$
\end{enumerate}
Since we have constructed a non--commutative version of the Veronese mapping, 
a natural question arises : What are non--commutative versions 
corresponding to 1. and 2. above ?

\par \noindent
These are very interesting problems from the view point of non--commutative 
``differential" geometry. It is worth challenging.

\subsection{Non--Commutative Pseudo Veronese Mapping}

We make a review of the noncompact one of Veronese mapping which we call 
a pseudo Veronese mapping. First let us define the manifold ${\fukuso}Q^{n}$ 
which is not always well known.
\begin{equation}
{\fukuso}Q^{n}=\left\{Q\in M(n+1;\fukuso)\ |\ 
Q^{2}=Q,\ J_{n}QJ_{n}=Q^{\dagger}\ \mbox{and}\ \mbox{tr}Q=1\right\}
\end{equation}
where $J_{n}$ is a matrix defined by
\[
J_{n}=
\left(
  \begin{array}{ccccc}
    1 &    &       &       &           \\
      & -1 &       &       &           \\
      &    & \cdot &       &           \\
      &    &       & \cdot &           \\
      &    &       &       & (-1)^{n}
  \end{array}
\right)
\quad \in \quad  M(n+1;\fukuso).
\]
We note that this $J_{n}$ is not a convensional one. Usually it is taken as
\[
\widetilde{J}_{n}=
\left(
  \begin{array}{ccccc}
    1 &    &       &       &      \\
      & -1 &       &       &      \\
      &    & \cdot &       &      \\
      &    &       & \cdot &      \\
      &    &       &       & -1
  \end{array}
\right)
\quad \in \quad  M(n+1;\fukuso).
\]

For the space $H_{\fukuso}^{n}$ defined by
\begin{equation}
H_{\fukuso}^{n}=\left\{v\in \fukuso^{n+1}\ |\ v^{\dagger}J_{n}v=1\right\}
\end{equation}
we can define a map 
\begin{equation}
H_{\fukuso}^{n}\ \longrightarrow\ {\fukuso}Q^{n}\ :\ 
v\ \longmapsto\ Q=vv^{\dagger}J_{n}.
\end{equation}

For $Q_{1}\in {\fukuso}Q^{1}$ it can be written as 
\begin{eqnarray*}
Q_{1}&=&
\left(
  \begin{array}{cc}
   \alpha & -\bar{\beta}  \\
  -\beta  & \bar{\alpha}
  \end{array}
\right)
\left( 
  \begin{array}{ccccc}
    1 &    \\
      & 0  
  \end{array}
\right)
\left(
  \begin{array}{cc}
   \alpha & -\bar{\beta}  \\
  -\beta   & \bar{\alpha}
  \end{array}
\right)^{-1},
\quad | \alpha|^{2}-|\beta|^{2}=1      \\
&=&
v_{1}v_{1}^{\dagger}J_{1},
\end{eqnarray*}
where  
\[
v_{1}=  
\left(
  \begin{array}{c}
  \alpha \\
  -\beta
  \end{array}
\right)\ \in\ H_{\fukuso}^{1}.
\]
For this $v_{1}$ we define $v_{n}$ as
\begin{equation}
\label{eq:map noncompact}
v_{n}=
\left(
  \begin{array}{c}
  \alpha^{n} \\
  \sqrt{{}_nC_1}\alpha^{n-1}(-\beta) \\
  \vdots \\
  \sqrt{{}_nC_j}\alpha^{n-j}(-\beta)^{j} \\
  \vdots \\
  (-\beta)^{n}
  \end{array}
\right).
\end{equation}
Then it is easy to see
\[
v_{n}^{\dagger}J_{n}v_{n}=\left(|\alpha|^{2}-|\beta|^{2}\right)^{n}=1,
\]
so $v_{n}\in H_{\fukuso}^{n}$. Namely, we defined the map 
\[
H_{\fukuso}^{1}\ \longrightarrow\ H_{\fukuso}^{n}\ :\ 
v_{1}\ \longmapsto\ v_{n}. 
\]
Therefore, we have the noncompact one of Veronese mapping
\begin{equation}
{\fukuso}Q^{1} \longrightarrow {\fukuso}Q^{n}\ :\ 
Q_{1}=v_{1}v_{1}^{\dagger}J_{1}\ \longmapsto\ 
Q_{n}=v_{n}v_{n}^{\dagger}J_{n}.
\end{equation}
For a tentative terminology let us call this a pseudo Veronese mapping.

Next, let us consider a local coordinate system. 
From (\ref{eq:map noncompact})
\[
v_{n}=
\left(
  \begin{array}{c}
  1 \\
  \sqrt{{}_nC_1}w \\
  \vdots \\
  \sqrt{{}_nC_j}w^{j} \\
  \vdots \\
  w^{n}
  \end{array}
\right)\alpha^{n}
\quad \mbox{where}\quad w=-\alpha/\beta
\]
then it is easy to check $|w|^{2} < 1$. We define a domain like 
(open) hyperbolic pillar
\[
D_{J}^{n}=\left\{v\in \fukuso^{n}\ |\ v^{\dagger}J_{n-1}v < 1\right\}.
\]
Then 
\[
D_{J}^{1}\ \longrightarrow\ D_{J}^{n}\ :\ 
w \longmapsto 
\left(
  \begin{array}{c}
  \sqrt{{}_nC_1}w \\
  \vdots \\
  \sqrt{{}_nC_j}w^{j} \\
  \vdots \\
  w^{n}
  \end{array}
\right)
\]
is a local map that we are looking for. As a whole see the following picture.

\vspace{-5mm}
\setlength{\unitlength}{1mm} 
\begin{center}
\begin{picture}(70,60)
\put(20,26){\vector(1,0){30}}
\put(20,50){\vector(1,0){30}}
\put(20, 2){\vector(1,0){30}}
\put(12,45){\vector(0,-1){15}}
\put(57,45){\vector(0,-1){15}}
\put(10,23){\makebox(6,6)[c]{${\fukuso}Q^{1}$}}
\put(55,23){\makebox(6,6)[c]{${\fukuso}Q^{n}$}}
\put(10,47){\makebox(6,6)[c]{$H_{\fukuso}^{1}$}}
\put(55,47){\makebox(6,6)[c]{$H_{\fukuso}^{n}$}}
\put(10,-1){\makebox(6,6)[c]{${D_{J}}^{1}$}}
\put(55,-1){\makebox(6,6)[c]{${D_{J}}^{n}$}}
\put(12, 6){\vector(0, 1){15}}
\put(57, 6){\vector(0, 1){15}}
\end{picture}
\end{center}

Next we want to consider a non--commutative version of the map. 
If we set 
\begin{equation}
{\cal B}\equiv 
\left(
  \begin{array}{c}
   \Gamma_{0} \\
   \Omega_{0}
  \end{array}
\right)
=
\left(
  \begin{array}{c}
    \frac{S(N+1)+\theta}{\sqrt{2S(N+1)(S(N+1)+\theta)}} \\
   -\frac{1}{\sqrt{2S(N)(S(N)+\theta)}}a^{\dagger}
  \end{array}
\right)
\end{equation}
from $V$ in (\ref{eq:W-I}), then 
\[
{\cal B}^{\dagger}{\bf J}{\cal B}=\Gamma_{0}^{2}-
\Omega_{0}^{\dagger}\Omega_{0}={\bf 1}
\quad \mbox{and}\quad 
\Omega_{0}\Gamma_{0}^{-1}=\frac{-1}{S(N)+\theta}a^{\dagger}\equiv W.
\]
That is, ${\cal B}=(\Gamma_{0},\Omega_{0})^{T}$ is a non--commutative 
hyperboloid and $W$ is a kind of ``stereographic projection" of 
the hyperboloid. 
It is easy to see the following 
\begin{equation}
\label{eq:easy-formula noncompact}
{\bf 1}-W^{\dagger}W=\frac{2S(N+1)}{S(N+1)+\theta}=\Gamma_{0}^{-2}
\Longrightarrow 
\Gamma_{0}=\left({\bf 1}-W^{\dagger}W\right)^{-1/2}.
\end{equation}

Here let us introduce new notations for the following. For $j\geq 0$ we set 
\begin{eqnarray}
\label{eq:def-gamma}
\Gamma_{-j}&=&\frac{S(N+1-j)+\theta}{\sqrt{2S(N+1-j)(S(N+1-j)+\theta)}}, \\
\label{eq:def-omega}
\Omega_{-j}&=-&\sqrt{\frac{N-j}{N}}\frac{1}{\sqrt{2S(N-j)(S(N-j)+\theta)}}
a^{\dagger}.
\end{eqnarray}
We list some useful formulas.
\begin{equation}
\label{eq:useful-formulas noncompact}
\Gamma_{-j}^{2}-\Omega_{-j}^{\dagger}\Omega_{-j}={\bf 1}
\quad \mbox{and}\quad
\Omega_{-j}^{\dagger}\Omega_{-j}=
\Omega_{-j+1}\Omega_{-j+1}^{\dagger}\quad 
\mbox{for}\quad j\geq 0.
\end{equation}

Now we are in a position to define a non--commutative version of the pseudo 
Veronese mapping.
\begin{equation}
\label{eq:Quantum-Mapping noncompact}
{\cal B}=
\left(
  \begin{array}{c}
   \Gamma_{0} \\
   \Omega_{0}
  \end{array}
\right)
\longrightarrow 
{\cal B}_{n}=
\left(
  \begin{array}{c}
   \Gamma_{0}^{n} \\
   \sqrt{{}_nC_1}\Omega_{0}\Gamma_{0}^{n-1} \\
   \vdots \\
   \sqrt{{}_nC_j}\Omega_{-(j-1)}\Omega_{-(j-2)}\cdots \Omega_{-1}\Omega_{0}
   \Gamma_{0}^{n-j} \\
   \vdots \\
   \sqrt{{}_nC_{n-1}}\Omega_{-(n-2)}\Omega_{-(n-3)}\cdots \Omega_{-1}
   \Omega_{0}\Gamma_{0} \\
   \Omega_{-(n-1)}\Omega_{-(n-2)}\Omega_{-(n-3)}\cdots \Omega_{-1}\Omega_{0}
  \end{array}
\right).
\end{equation}
Then it is not difficult to see 
\[
{\cal B}_{n}^{\dagger}{\bf J}_{n}{\cal B}_{n}
=\left(\Gamma_{0}^{2}-\Omega_{0}^{\dagger}\Omega_{0}\right)^{n}
=\left({\cal B}^{\dagger}{\bf J}{\cal B}\right)^{n}
={\bf 1},
\]
where ${\bf J}_{n}$ (${\bf J}_{1}={\bf J}$) is defined by
\[
{\bf J}_{n}=
\left(
  \begin{array}{ccccc}
   {\bf 1} &          &       &       &                 \\
           & -{\bf 1} &       &       &                 \\
           &          & \cdot &       &                 \\
           &          &       & \cdot &                 \\
           &          &       &       & (-1)^{n}{\bf 1}
  \end{array}
\right)
=J_{n}\otimes {\bf 1}.
\]

From this we can define the projectors which correspond to pseudo projective 
spaces like 
\begin{equation}
{\cal Q}_{n}={\cal B}_{n}{\cal B}_{n}^{\dagger}{\bf J}_{n},
\quad 
{\cal Q}_{1}={\cal B}{\cal B}^{\dagger}{\bf J},
\end{equation}
so the map 
\begin{equation}
\label{eq:Quantum-Veronese-Mapping noncompact}
{\cal Q}_{1}\ \longrightarrow \ {\cal Q}_{n}
\end{equation}
is a non-commutative version of the pseudo Veronese mapping.

\section{Non--Commutative Representation Theory}

In this section we construct a map (in the non--commutative models) 
corresponding to spin $j$--representation for the compact group $SU(2)$ and 
noncompact group $SU(1,1)$$\cdots$ a kind of non--commutative version of 
classical spin representations $\cdots$.

\subsection{Non--Commutative Version of $SU(2)$ Case}

The construction of spin $j$--representation ($j \in \seisu_{\geq 0}+1/2$) 
is well--known. Let us make a brief review within our necessity. 
For the vector space 
\[
{\cal H}_{J}=\mbox{Vect}_{\fukuso}\left\{{\sqrt{{}_{J-1}C_k}z^{k}}\ 
\vert \ 
k \in \{0,1,\cdots,J-1\}\right\}
\]
where $J=2j+1 (\in \futon)$, the inner product in this space is given by
\[
<f|g>=\frac{2J}{2\pi}\int_{\fukuso}
\frac{d^{2}z}{(1+|z|^{2})^{J+1}}f(z)\overline{g(z)}
=\sum_{k=0}^{J-1}a_{k}\bar{b}_{k}
\]
for $f(z)=\sum_{k=0}^{J-1}\sqrt{{}_{J-1}C_k}a_{k}z^{k}$ and 
$g(z)=\sum_{k=0}^{J-1}\sqrt{{}_{J-1}C_k}b_{k}z^{k}$ in ${\cal H}_{J}$. 
Here $d^{2}z$ means $dxdy$ for $z=x+iy$.

\par \noindent
For example, for $j=1/2$, $j=1$ and $j=3/2$ 
\[
{\cal H}_{2}=\mbox{Vect}_{\fukuso}\left\{1,\ z\right\},\quad 
{\cal H}_{3}=\mbox{Vect}_{\fukuso}\left\{1,\ \sqrt{2}z,\ z^{2}\right\},\quad 
{\cal H}_{4}=\mbox{Vect}_{\fukuso}\left\{1,\ \sqrt{3}z,\ \sqrt{3}z^{2},\ 
z^{3}\right\}.
\]
Therefore, we identify ${\cal H}_{J}$ with $\fukuso^{J}$ by
\[
f(z)=\sum_{k=0}^{J-1}\sqrt{{}_{J-1}C_k}a_{k}z^{k}\ \longleftrightarrow\ 
(a_{0},a_{1},\cdots,a_{J-1})^{\mbox{T}}.
\]

For 
\[
A=
\left(
  \begin{array}{cc}
    \alpha & -\bar{\beta} \\
    \beta  &  \bar{\alpha}
  \end{array}
\right)\ \in SU(2)\quad \left(|\alpha|^{2}+|\beta|^{2}=1\right)
\]
the spin $j$ representation 
\[
\phi_{j} : SU(2) \longrightarrow SU(J)
\]
is defined as 
\begin{equation}
\label{eq:spin-representation}
\left(\phi_{j}(A)f\right)(z)=(\alpha+\beta z)^{J-1}
f\left(\frac{-\bar{\beta}+\bar{\alpha}z}{\alpha+\beta z}\right)
\end{equation}
where $f \in {\cal H}_{J}$. It is easy to obtain $\phi_{j}(A)$ for $j=1/2$, 
$j=1$ and $j=3/2$.

Namely, the spin 1/2 representation is 
\begin{equation}
\label{eq:spin-1/2}
\phi_{1/2}(A)=
\left(
  \begin{array}{cc}
    \alpha & -\bar{\beta} \\
    \beta  &  \bar{\alpha}
  \end{array}
\right)=A,
\end{equation}
the spin 1 representation is 
\begin{equation}
\label{eq:spin-1}
\phi_{1}(A)=
\left(
  \begin{array}{ccc}
    \alpha^{2} & -\sqrt{2}\alpha\bar{\beta} & \bar{\beta}^{2} \\
    \sqrt{2}\alpha\beta & |\alpha|^{2}-|\beta|^{2} & 
    -\sqrt{2}\bar{\alpha}\bar{\beta} \\
    \beta^{2} & \sqrt{2}\bar{\alpha}\beta & \bar{\alpha}^{2}
  \end{array}
\right),
\end{equation}
and the spin 3/2 representation is 
\begin{equation}
\label{eq:spin-3/2}
\phi_{3/2}(A)=
\left(
  \begin{array}{cccc}
    \alpha^{3} & -\sqrt{3}\alpha^{2}\bar{\beta} & 
    \sqrt{3}\alpha\bar{\beta}^{2} & -\bar{\beta}^{3} \\
    \sqrt{3}\alpha^{2}\beta & (|\alpha|^{2}-2|\beta|^{2})\alpha & 
    -(2|\alpha|^{2}-|\beta|^{2})\bar{\beta} & 
    \sqrt{3}\bar{\alpha}\bar{\beta}^{2} \\
    \sqrt{3}\alpha\beta^{2} & (2|\alpha|^{2}-|\beta|^{2})\beta & 
    (|\alpha|^{2}-2|\beta|^{2})\bar{\alpha} & 
    -\sqrt{3}\bar{\alpha}^{2}\bar{\beta} \\
    \beta^{3} & \sqrt{3}\bar{\alpha}\beta^{2} & 
    \sqrt{3}\bar{\alpha}^{2}\beta & \bar{\alpha}^{3}
  \end{array}
\right).
\end{equation}

\par \vspace{5mm}
Next we want to consider a non--commutative version of the spin 
representation. However, since such a theory has not been known as far as 
we know we must look for mappings corresponding to $\phi_{1}(A)$ and 
$\phi_{3/2}(A)$ by (many) trial and error, see Appendix C.

If we set 
\[
U\equiv U_{I}=
\left(
  \begin{array}{cc}
    X_{0} & -Y_{0}^{\dagger} \\
    Y_{0} &  X_{-1}
  \end{array}
\right)\ :\ \mbox{unitary}
\]
from (\ref{eq:V-I}), then the corresponding map for $\phi_{1}(A)$ is 
\begin{equation}
\label{eq:non-spin-1}
\Phi_{1}(U)=
\left(
 \begin{array}{ccc}
  X_{0}^{2}& -\sqrt{2}X_{0}Y_{0}^{\dagger}& Y_{0}^{\dagger}Y_{-1}^{\dagger} \\
  \sqrt{2}Y_{0}X_{0}& X_{-1}^{2}-Y_{-1}^{\dagger}Y_{-1}& 
  -\sqrt{2}X_{-1}Y_{-1}^{\dagger} \\
  Y_{-1}Y_{0}& \sqrt{2}Y_{-1}X_{-1}& X_{-2}^{2}
  \end{array}
\right)
\end{equation}             
and the corresponding map for $\phi_{3/2}(A)$ is 
\begin{eqnarray}
\label{eq:non-spin-3/2}
\Phi_{3/2}(U)&=&
\left(
 \begin{array}{cccc}
  X_{0}^{3}& 
  -\sqrt{3}X_{0}^{2}Y_{0}^{\dagger}& 
  \sqrt{3}X_{0}Y_{0}^{\dagger}Y_{-1}^{\dagger}& 
  -Y_{0}^{\dagger}Y_{-1}^{\dagger}Y_{-2}^{\dagger} \\
  \sqrt{3}Y_{0}X_{0}^{2}& 
  X_{-1}\left(X_{-1}^{2}-2Y_{-1}^{\dagger}Y_{-1}\right)&
  -\left(2X_{-1}^{2}-Y_{-1}^{\dagger}Y_{-1}\right)Y_{-1}^{\dagger}&
  \sqrt{3}X_{-1}Y_{-1}^{\dagger}Y_{-2}^{\dagger} \\
  \sqrt{3}Y_{-1}Y_{0}X_{0}&
  Y_{-1}\left(2X_{-1}^{2}-Y_{-1}^{\dagger}Y_{-1}\right)&
  X_{-2}\left(X_{-2}^{2}-2Y_{-2}^{\dagger}Y_{-2}\right)&
  -\sqrt{3}X_{-2}^{2}Y_{-2}^{\dagger} \\
  Y_{-2}Y_{-1}Y_{0}&
  \sqrt{3}Y_{-2}Y_{-1}X_{-1}&
  \sqrt{3}Y_{-2}X_{-2}^{2}&
  X_{-3}^{3}
  \end{array}
\right).      \nonumber \\
&&{}
\end{eqnarray}
To check the unitarity of $\Phi_{1}(U)$ and $\Phi_{3/2}(U)$ is long but 
straightforward. 

For $j\geq 2$ we could not find a general method like 
(\ref{eq:spin-representation}) which determines $\Phi_{j}(U)$. 
However, we know only that the first column of $\Phi_{j}(U)$ is just 
${\cal A}_{2j}$ in (\ref{eq:Quantum-Mapping}).,
\[
\Phi_{j}(U)=\left({\cal A}_{2j},*,\cdots,*\right)\ :\ \mbox{unitary}
\]
and 
\[
\Phi_{j}(U)
\left(
 \begin{array}{ccccc}
 {\bf 1} &         &       &       &         \\
         & {\bf 0} &       &       &         \\
         &         & \cdot &       &         \\
         &         &       & \cdot &         \\
         &         &       &       & {\bf 0} \\
  \end{array}
\right)
\Phi_{j}(U)^{\dagger}
=
{\cal A}_{2j}{\cal A}_{2j}^{\dagger}
=
{\cal P}_{2j}.
\]

\par \noindent
We leave finding a general method to the readers as a challenging problem.

\subsection{Non--Commutative Version of $SU(1,1)$ Case}

Let us review some aspects of the theory of unitary representation of 
$SU(1,1)$ within our necessity. 

Let $H^{2}\equiv H^{2}(D)$ be the second 
Hardy class where $D$ is the open unit disk in $\fukuso$. We consider the 
spin $j$ representation of the non--compact group $SU(1,1)$.
The inner product is defined as 
\[
<f|g>
=\frac{2(2j-1)}{2\pi}\int_{D} d^{2}z(1-|z|^{2})^{2j-2}f(z)\overline{g(z)}
=\sum_{n=0}^{\infty}\frac{n!}{(2j)_{n}}a_{n}\bar{b}_{n}
\]
for $f(z)=\sum_{n=0}^{\infty}a_{n}z^{n}$ and 
$g(z)=\sum_{n=0}^{\infty}b_{n}z^{n}$ in $H^{2}$. For $j=\frac{1}{2}$ we must 
take some renormalization into consideration (we omit it here). 
Then $\left\{H^{2},<|>\right\}$ becomes a complex Hibert space. 

Therefore, it is better for us to consider the vector space 
\[
H^{2}_{2j}=\mbox{Vect}_{\fukuso}\left\{1,\sqrt{2j}z,\cdots,
\sqrt{\frac{(2j)_{k}}{k!}}z^{k},\cdots \right\}
\]
and the correspondence between $H^{2}_{2j}$ and $\ell^{2}(\fukuso)$ is given 
by 
\[
f(z)=\sum_{n=0}^{\infty}\sqrt{\frac{(2j)_{n}}{n!}}a_{n}z^{n}\ 
\longleftrightarrow\ 
(a_{0},a_{1},\cdots,a_{n},\cdots)^{\mbox{T}},
\]
so we identify $H^{2}_{2j}$ with $\ell^{2}(\fukuso)$ by this correspondence.

For 
\[
B=
\left(
  \begin{array}{cc}
    \alpha & -\bar{\beta}  \\
    -\beta &  \bar{\alpha}
  \end{array}
\right)\ \in SU(1,1)\quad \left(|\alpha|^{2}-|\beta|^{2}=1\right)
\]
the spin $j$ unitary representation 
\[
\psi_{j} : SU(1,1) \longrightarrow U(\ell^{2}(\fukuso))
\]
is defined as 
\begin{equation}
\label{eq:spin-representation noncompact}
\left(\psi_{j}(B)f\right)(z)=(\alpha+\beta z)^{-2j}
f\left(\frac{\bar{\beta}+\bar{\alpha}z}{\alpha+\beta z}\right)
\end{equation}
where $f \in H^{2}_{2j}$. 

\par \noindent 
For example, when $f=1$ (a constant) it is easy to see
\begin{eqnarray*}
\left(\psi_{j}(B)1\right)(z)
&=&(\alpha+\beta z)^{-2j}         \\
&=&\frac{1}{\alpha^{2j}}-2j\frac{\beta}{\alpha^{2j+1}}z+\cdots+
(-1)^{n}\frac{(2j)_{n}}{n!}\frac{\beta^{n}}{\alpha^{2j+n}}z^{n}+\cdots  \\
&=&\frac{1}{\alpha^{2j}}-\sqrt{2j}\frac{\beta}{\alpha^{2j+1}}\sqrt{2j}z+
\cdots+
(-1)^{n}\sqrt{\frac{(2j)_{n}}{n!}}\frac{\beta^{n}}{\alpha^{2j+n}}
\sqrt{\frac{(2j)_{n}}{n!}}z^{n}+\cdots
\end{eqnarray*}
where $(a)_{n}$ is the Pochammer notation defined by 
\[
(a)_{n}=a(a+1)\cdots(a+n-1).
\]
Therefore 
\begin{equation}
\label{eq:first-column}
\psi_{j}(B)1
=
\left(
  \begin{array}{c}
  \frac{1}{\alpha^{2j}} \\
  -\sqrt{2j}\frac{\beta}{\alpha^{2j+1}} \\
  \vdots \\
  (-1)^{n}\sqrt{\frac{(2j)_{n}}{n!}}\frac{\beta^{n}}{\alpha^{2j+n}} \\
  \vdots \\
  \end{array}
\right)
=
\left(
  \begin{array}{c}
  \alpha^{-2j} \\
  -\sqrt{2j}{\beta}\alpha^{-(2j+1)} \\
  \vdots \\
  (-1)^{n}\sqrt{\frac{(2j)_{n}}{n!}}\beta^{n}\alpha^{-(2j+n)} \\
  \vdots \\
  \end{array}
\right).
\end{equation}

More generally, for $f_{k}(z)=\sqrt{\frac{(2j)_{k}}{k!}}z^{k}$
\begin{eqnarray*}
\left(\psi_{j}(B)f_{k}\right)(z)
&=&\sqrt{\frac{(2j)_{k}}{k!}}
(\bar{\beta}+\bar{\alpha}z)^{k}(\alpha+\beta z)^{-(2j+k)}   \\
&=&\sqrt{\frac{(2j)_{k}}{k!}}\sum_{l=0}^{k} \sum_{n=0}^{\infty}
{{}_kC_l}(-1)^{n}\frac{(2j+k)_{n}}{n!}
\frac{\beta^{n}\bar{\beta}^{k-l}\bar{\alpha}^{l}}{\alpha^{2j+n+k}}
z^{l+n},
\end{eqnarray*}
so
\[
\psi_{j}(B)f_{k}
=
\left(
  \begin{array}{c}
   \sqrt{\frac{(2j)_{k}}{k!}}\frac{\bar{\beta}^{k}}{\alpha^{2j+k}}   \\
   \sqrt{\frac{(2j)_{k}}{k!}}\sqrt{\frac{1}{2j}}
   \left\{k|\alpha|^{2}-(2j+k)|\beta|^{2}\right\}
   \frac{\bar{\beta}^{k-1}}{\alpha^{2j+k+1}}                         \\
   \sqrt{\frac{(2j)_{k}}{k!}}\sqrt{\frac{2!}{(2j)_{2}}}
   \left\{\frac{k(k-1)}{2}|\alpha|^{4}-k(2j+k)|\alpha|^{2}|\beta|^{2}+
   \frac{(2j+k)(2j+k+1)}{2}|\beta|^{4}\right\}
   \frac{\bar{\beta}^{k-2}}{\alpha^{2j+k+2}}                         \\
  \vdots \\
  \vdots
  \end{array}
\right).
\]
Therefore the matrix defined by
\begin{equation}
\psi_{j}(B)=\left(\psi_{j}(B)f_{0},\psi_{j}(B)f_{1},\cdots,\psi_{j}(B)f_{k},
\cdots\right)
\end{equation}
is the unitary representation that we are looking for.

We set
\[
V=
\left(
  \begin{array}{cc}
    \Gamma_{0} & \Omega_{0}^{\dagger} \\
    \Omega_{0} & \Gamma_{-1}
  \end{array}
\right)\ :\ \mbox{pseudo unitary}\ (V^{\dagger}{\bf J}V={\bf J})
\]
from (\ref{eq:W-I}). In this case it is almost impossible to obtain 
the explicit unitary operator $\Psi_{j}(V)$ corresponding to $\psi_{j}(B)$.

However, we can at least determine the first column of $\Psi_{j}(V)$ by 
making use of (\ref{eq:first-column}) : 
\begin{equation}
\widehat{\cal B}\equiv 
\left(
  \begin{array}{c}
  \Gamma_{0}^{-2j} \\
  \sqrt{2j}{\Omega_{0}}\Gamma_{0}^{-(2j+1)} \\
  \vdots \\
  \sqrt{\frac{(2j)_{n}}{n!}}
  \Omega_{-(n-1)}\Omega_{-(n-2)}\Omega_{-(n-3)}\cdots \Omega_{-1}\Omega_{0}
  \Gamma_{0}^{-(2j+n)} \\
  \vdots \\
  \end{array}
\right)
\end{equation}
with $\Gamma_{0}$ and $\Omega_{-j}$ in (\ref{eq:def-gamma}) and 
(\ref{eq:def-omega}). Then it is not difficult to see 
\[
\widehat{\cal B}^{\dagger}\widehat{\cal B}
=
\left(\Gamma_{0}^{2}-\Omega_{0}^{\dagger}\Omega_{0}\right)^{-2j}
=
{\bf 1}.
\]
Therefore, the map making use of projectors 
\begin{equation}
{\cal Q}_{1}={\cal B}{\cal B}^{\dagger}{\bf J}\quad \longrightarrow \quad 
\widehat{\cal P}=\widehat{\cal B}\widehat{\cal B}^{\dagger}
\end{equation}
is a non--commutative version of the unitary expression of pseudo 
Veronese mapping. Compare the discussion here with the one after 
the equation (\ref{eq:Quantum-Mapping noncompact}). 

We note that if the unitary operator
\[
\Psi_{j}(V)=\left(\widehat{\cal B}_{0},\widehat{\cal B}_{1},\cdots,
\widehat{\cal B}_{n},\cdots\right),\quad \widehat{\cal B}_{0}=\widehat{\cal B}
\]
could be defined (we cannot determine $\widehat{\cal B}_{n}$ for $n\geq 1$), 
then we have
\[
\Psi_{j}(V)
\left(
 \begin{array}{cccc}
 {\bf 1} &         &         &        \\
         & {\bf 0} &         &        \\
         &         & {\bf 0} &        \\
         &         &         & \ddots
  \end{array}
\right)
\Psi_{j}(V)^{\dagger}
=\widehat{\cal B}\widehat{\cal B}^{\dagger}
=\widehat{\cal P}.
\]

\par
A comment is in order.\ In the construction of $\widehat{\cal B}_{n}$ we need 
an infinite number of operators, which means a kind of 
{\bf second non--commutativization}.

\section{Discussion}

In this paper we derived a non--commutative version of the Berry model 
(based on $SU(2)$) arising from the Jaynes--Cummings model in quantum optics 
and the pseudo Berry model (based on $SU(1,1)$) by changing the generators, 
and constructed a non--commutative version of the Hopf and pseudo Hopf bundles 
in the classical case. 

The bundle has a kind of Dirac strings in the case of non--commutative Berry 
model. However, they appear in only states containing the ground one 
(${\cal F}\times \{\ket{0}\} \cup \{\ket{0}\}\times {\cal F} \subset 
{\cal F}\times {\cal F}$) 
and don't appear in excited states, which is very interesting. 

In general, a non-commutative version of classical field theory is of course 
not unique. If our model is a ``correct" one, then this paper give an example 
that classical singularities like Dirac strings are not universal in some 
non--commutative model. 
As to general case with higher spins which are not easy, see \cite{TS}.

Moreover, for the two models a non--commutative version of the Veronese 
mapping or pseudo Veronese mapping was constructed, and unitary mappings 
corresponding to (classical) spin representations were constructed though 
they are not necessarily enough.

The results or methods in the paper will become a starting point to construct 
a fruitful non--commutative geometry or representation theory.

Last, we would like to make a comment. To develop a ``quantum" mathematics 
we need a rigorous method to treat an analysis or a geometry on infinite 
dimensional spaces like Fock space. 
In quantum field theories physicists have given some (interesting) methods, 
while they are more or less formal from the mathematical point of view. 
It is a rigorous method which we need. As a trial \cite{Asada} is recommended.

\vspace{5mm}
\noindent{\em Acknowledgment.}\\
The author wishes to thank Akira Asada, Yoshinori Machida, Shin'ichi Nojiri, 
Ryu Sasaki and Tatsuo Suzuki for their helpful comments and suggestions. 

\par \noindent
The author also thanks to Gennadi Sardanashvily and Giovanni Giachetta for 
warm hospitality at Firenze (14-18/April/2005). The arrangement of this paper 
was determined during the stay.

\par \vspace{3mm}
\begin{center}
 \begin{Large}
   {\bf Appendix}
 \end{Large}
\end{center}

\vspace{5mm} \noindent
\begin{Large}
{\bf A\ \ Classical Theory of Projective Spaces}
\end{Large}

\par \vspace{3mm} \noindent
Complex projective spaces are typical examples of symmetric spaces and are 
very tractable, so they are used to construct several examples in both 
physics and mathematics. 
We make a review of complex projective spaces within our necessity, see for 
example \cite{MN}, \cite{Fujii}, \cite{FKSF}. 

For $n \in \futon$ the complex projective space ${\fukuso}P^{n}$ is defined 
as follows : 
For \mbox{\boldmath $\zeta$}, \mbox{\boldmath $\mu$} $\in 
{\fukuso}^{n+1}-\{{\bf 0}\}$  \mbox{\boldmath $\zeta$} is equivalent to 
\mbox{\boldmath $\mu$} (\mbox{\boldmath $\zeta$} $\sim$ \mbox{\boldmath $\mu$}) if and only if 
\mbox{\boldmath $\zeta$} = $\lambda$\mbox{\boldmath $\mu$} for 
$\lambda \in \fukuso - \{0 \}$. 
We show the equivalent relation class as [\mbox{\boldmath $\zeta$}] and set 
${\fukuso}P^{n} \equiv {\fukuso}^{n+1}-\{{\bf 0}\} / \sim $. 
For   
\mbox{\boldmath $\zeta$} = $({\zeta}_{0}, {\zeta}_{1}, \cdots, {\zeta}_{n})$ 
we write usually as [\mbox{\boldmath $\zeta$}] = $[{\zeta}_{0}: {\zeta}_{1}:  
\cdots : {\zeta}_{n}]$. Then it is well--known that ${\fukuso}P^{n}$ has 
$n+1$ local charts, namely
\begin{equation}
  {\fukuso}P^{n} = \bigcup_{j=0}^{n} U_{j}\ ,  \quad 
    U_{j} = \{ [{\zeta}_{0}: \cdots : {\zeta}_{j}: \cdots : {\zeta}_{n}]\ |\  
          {\zeta}_{j} \ne 0 \}.
\end{equation} 
Since
\[
  ({\zeta}_{0}, \cdots , {\zeta}_{j}, \cdots , {\zeta}_{n}) =  
  {\zeta}_{j}\left(\frac{{\zeta}_{0}}{{\zeta}_{j}}, \cdots, 
  \frac{{\zeta}_{j-1}}{{\zeta}_{j}}, 1, \frac{{\zeta}_{j+1}}{{\zeta}_{j}}, 
  \cdots, \frac{{\zeta}_{n}}{{\zeta}_{j}}\right),
\]
we have the local coordinate on $U_{j}$ 
\begin{equation}
  \left(\frac{{\zeta}_{0}}{{\zeta}_{j}}, \cdots, 
  \frac{{\zeta}_{j-1}}{{\zeta}_{j}}, \frac{{\zeta}_{j+1}}{{\zeta}_{j}}, 
  \cdots, \frac{{\zeta}_{n}}{{\zeta}_{j}}\right). 
\end{equation}

However the above definition of ${\fukuso}P^{n}$ is not tractable, so we use 
the well--known expression by projections (see \cite{Fujii})
\begin{equation}
 {\fukuso}P^{n} \cong G_{1,n+1}(\fukuso) = 
     \{P \in M(n+1; \fukuso)\ |\ P^{2} = P,\ P = P^{\dagger}\ \mbox{and}\ 
       \mbox{tr}P = 1 \}
\end{equation}
and the correspondence 
\begin{equation}
 \label{eq:correspondence}
  [{\zeta}_{0}: {\zeta}_{1}: \cdots : {\zeta}_{n}] \Longleftrightarrow 
  \frac{1}{\zetta{{\zeta}_{0}}^2 + \zetta{{\zeta}_{1}}^2 + \cdots + 
          \zetta{{\zeta}_{n}}^2 }
  \left(
     \begin{array}{ccccc} 
         \zetta{{\zeta}_{0}}^2& {\zeta}_{0}{\bar {\zeta}_{1}}& 
         \cdot& \cdot& {\zeta}_{0}{\bar {\zeta}_{n}}  \\
         {\zeta}_{1}{\bar {\zeta}_{0}} & \zetta{{\zeta}_{1}}^2&  
         \cdot& \cdot& {\zeta}_{1}{\bar {\zeta}_{n}}  \\
         \cdot& \cdot& & & \cdot  \\
         \cdot& \cdot& & & \cdot  \\
         {\zeta}_{n}{\bar {\zeta}_{0}}& {\zeta}_{n}{\bar {\zeta}_{1}}& 
         \cdot& \cdot& \zetta{{\zeta}_{n}}^2     
     \end{array}
  \right) \equiv P\ .
\end{equation}
If we set 
\begin{equation}
  \ket{\mbox{\boldmath $\zeta$}}=
 \frac{1}{\sqrt{\sum_{j=0}^{n} \zetta{\zeta_{j}}^2} }
  \left(
     \begin{array}{c}
        {\zeta}_{0} \\
        {\zeta}_{1} \\
         \cdot  \\
         \cdot  \\
        {\zeta}_{n} 
     \end{array} 
  \right)\ , 
\end{equation}
then we can write the right hand side of (\ref{eq:correspondence}) as 
\begin{equation}
 \label{eq:projection}
  P = \ket{\mbox{\boldmath $\zeta$}}\bra{\mbox{\boldmath $\zeta$}} \quad 
  \mbox{and} \quad 
    \braket{\mbox{\boldmath $\zeta$}}{\mbox{\boldmath $\zeta$}} = 1.
\end{equation}
For example on $U_{0}$ 
\[
  \left(z_{1}, z_{2}, \cdots, z_{n} \right) = 
  \left(\frac{{\zeta}_{1}}{{\zeta}_{0}},\frac{{\zeta}_{2}}{{\zeta}_{0}}, 
  \cdots, 
  \frac{{\zeta}_{n}}{{\zeta}_{0}}\right) ,  
\]
we have 
\begin{eqnarray}
  P(z_{1}, \cdots, z_{n}) &=& 
  \frac{1}{1 + \sum_{j=1}^{n} \zetta{z_{j}}^2}
     \left(
         \begin{array}{ccccc}
             1& {\bar z_{1}}& \cdot& \cdot& {\bar z_{n}}  \\
             z_{1}& \zetta{z_{1}}^2& \cdot& \cdot& z_{1}{\bar z_{n}} \\
             \cdot& \cdot& & & \cdot \\
             \cdot& \cdot& & & \cdot \\
             z_{n}& z_{n}{\bar z_{1}}& \cdot& \cdot& \zetta{z_{n}}^2
         \end{array}
     \right)  \nonumber \\
   &=& \ket{\left(z_{1}, z_{2}, \cdots, z_{n}\right)}
       \bra{\left(z_{1}, z_{2}, \cdots, z_{n}\right)}\ ,
\end{eqnarray}
where 
\begin{equation}
  \ket{\left(z_{1}, z_{2}, \cdots, z_{n} \right)} = 
  \frac{1}{\sqrt{1 + \sum_{j=1}^{n} \zetta{z_{j}}^2}}
     \left(
         \begin{array}{c}
             1 \\
             z_{1} \\
             \cdot \\
             \cdot \\
             z_{n} 
         \end{array}
     \right).  \nonumber \\
\end{equation}

To be clearer, let us give a detailed description for the case of $n$ = $1$ 
and $2$.

\par \noindent
(a) {\bf $n = 1$} : 
\begin{eqnarray}
 \label{eq:cp1-1}
  P(z)&=&\frac{1}{1+\zetta{z}^2}
     \left(
         \begin{array}{cc}
             1& {\bar z} \\
             z& \zetta{z}^2 
         \end{array}
     \right)  
   = \ket{z}\bra{z}, \nonumber  \\
  &&\mbox{where}\ \ket{z}=\frac{1}{\sqrt{1+\zetta{z}^2}}
     \left(
         \begin{array}{c}
             1 \\
             z 
         \end{array}
     \right), 
  \quad z=\frac{\zeta_{1}}{\zeta_{0}}, \quad \mbox{on}\ U_{0}\ ,  \\
 \label{eq:cp1-2}
  P(w)&=&\frac{1}{\zetta{w}^2+1}
     \left(
         \begin{array}{cc}
             \zetta{w}^2 & w \\
             {\bar w}& 1
         \end{array}
     \right)  
   = \ket{w}\bra{w},  \nonumber  \\
  &&\mbox{where}\ \ket{w}=\frac{1}{\sqrt{\zetta{w}^2+1}}
     \left(
         \begin{array}{c}
             w \\
             1
         \end{array}
     \right), 
  \quad w=\frac{\zeta_{0}}{\zeta_{1}}, \quad \mbox{on}\ U_{1}\ . 
\end{eqnarray}

\vspace{5mm}
\par \noindent
(b) {\bf $n = 2$} : 
\begin{eqnarray}
 \label{eq:cp2-1}
  P(z_{1},z_{2})&=&\frac{1}{1+\zetta{z_{1}}^2+\zetta{z_{2}}^2}
     \left(
         \begin{array}{ccc}
             1& {\bar z_{1}}& {\bar z_{2}} \\
             z_{1}& \zetta{z_{1}}^2& z_{1}{\bar z_{2}} \\
             z_{2}& z_{2}{\bar z_{1}}& \zetta{z_{2}}^2 
         \end{array}
     \right)  
   = \ket{(z_{1},z_{2})}\bra{(z_{1},z_{2})}, \nonumber  \\
  \mbox{where} 
  &&\ket{(z_{1},z_{2})}=\frac{1}{\sqrt{1+\zetta{z_{1}}^2+\zetta{z_{2}}^2}}
     \left(
         \begin{array}{c}
             1 \\
             z_{1} \\
             z_{2} 
         \end{array}
     \right), 
\quad (z_{1},z_{2})=\left(\frac{\zeta_{1}}{\zeta_{0}},
   \frac{\zeta_{2}}{\zeta_{0}} \right)  \quad \mbox{on}\ U_{0}\ ,  \\
 \label{eq:cp2-2}
  P(w_{1},w_{2})&=&\frac{1}{\zetta{w_{1}}^2+1+\zetta{w_{2}}^2}
     \left(
         \begin{array}{ccc}
             \zetta{w_{1}}^2& w_{1}& w_{1}{\bar w_{2}} \\
             {\bar w_{1}}& 1& {\bar w_{2}} \\
             w_{2}{\bar w_{1}}& w_{2}& \zetta{w_{2}}^2 
         \end{array}
     \right)  
   = \ket{(w_{1},w_{2})}\bra{(w_{1},w_{2})}, \nonumber \\
  \mbox{where} 
  &&\ket{(w_{1},w_{2})}=\frac{1}{\sqrt{\zetta{w_{1}}^2+1+\zetta{w_{2}}^2}}
     \left(
         \begin{array}{c}
             w_{1} \\
              1  \\
             w_{2} 
         \end{array}
     \right), 
\quad  (w_{1},w_{2})=\left(\frac{\zeta_{0}}{\zeta_{1}},
   \frac{\zeta_{2}}{\zeta_{1}} \right)\ \  \mbox{on}\ U_{1}\ , \\
 \label{eq:cp2-3}
  P(v_{1},v_{2})&=&\frac{1}{\zetta{v_{1}}^2+\zetta{v_{2}}^2+1}
     \left(
         \begin{array}{ccc}
             \zetta{v_{1}}^2& v_{1}{\bar v_{2}}& v_{1} \\
             v_{2}{\bar v_{1}}& \zetta{v_{2}}^2& v_{2} \\
             {\bar v_{1}}& {\bar v_{2}}& 1 
         \end{array}
     \right)  
   = \ket{(v_{1},v_{2})}\bra{(v_{1},v_{2})}, \nonumber \\
  \mbox{where} 
  &&\ket{(v_{1},v_{2})}=\frac{1}{\sqrt{\zetta{v_{1}}^2+\zetta{v_{2}}^2+1}}
     \left(
         \begin{array}{c}
             v_{1} \\
             v_{2}  \\
              1 
         \end{array}
     \right), 
\quad (v_{1},v_{2})=\left(\frac{\zeta_{0}}{\zeta_{2}},
   \frac{\zeta_{1}}{\zeta_{2}} \right)  \quad \mbox{on}\ U_{2}\ . 
\end{eqnarray}

\vspace{5mm} \noindent
\begin{Large}
{\bf B\ \ Local Coordinate of the Projector}
\end{Large}
\par \vspace{3mm} \noindent
We give a proof to the last formula in (\ref{eq:full-expression}). 

By making use of the expression by Oike in \cite{Fujii} (we don't 
repeat it here) 
\begin{equation}
\label{eq:oike-expression-1}
{\cal P}({\cal Z})=
\left(
  \begin{array}{cc}
    {\bf 1}  & -{\cal Z}^{\dagger} \\
    {\cal Z} & {\bf 1}
  \end{array}
\right)
\left(
  \begin{array}{cc}
    {\bf 1} &          \\
            & {\bf 0}
  \end{array}
\right)
\left(
  \begin{array}{cc}
    {\bf 1}  & -{\cal Z}^{\dagger} \\
    {\cal Z} & {\bf 1}
  \end{array}
\right)^{-1}
\end{equation}
where ${\cal Z}$ is some operator on the Fock space ${\cal F}$. 
Let us rewrite this into more useful form. From the simple relation 
\[
\left(
  \begin{array}{cc}
    {\bf 1}   & {\cal Z}^{\dagger} \\
    -{\cal Z} & {\bf 1}
  \end{array}
\right)
\left(
  \begin{array}{cc}
    {\bf 1}  & -{\cal Z}^{\dagger} \\
    {\cal Z} & {\bf 1}
  \end{array}
\right)
=
\left(
  \begin{array}{cc}
    {\bf 1}+{\cal Z}^{\dagger}{\cal Z}  &               \\
                  & {\bf 1}+{\cal Z}{\cal Z}^{\dagger}
  \end{array}
\right)
\]
we have 
\[
\left(
  \begin{array}{cc}
    {\bf 1}  & -{\cal Z}^{\dagger} \\
    {\cal Z} & {\bf 1}
  \end{array}
\right)^{-1}
=
\left(
  \begin{array}{cc}
    ({\bf 1}+{\cal Z}^{\dagger}{\cal Z})^{-1}  &          \\
             & ({\bf 1}+{\cal Z}{\cal Z}^{\dagger})^{-1}
  \end{array}
\right)
\left(
  \begin{array}{cc}
    {\bf 1}  & {\cal Z}^{\dagger} \\
   -{\cal Z} & {\bf 1}
  \end{array}
\right).
\]
Inserting this into (\ref{eq:oike-expression-1}) and some calculation leads to 
\begin{equation}
\label{eq:oike-expression-2}
{\cal P}({\cal Z})=
\left(
  \begin{array}{cc}
  ({\bf 1}+{\cal Z}^{\dagger}{\cal Z})^{-1} & 
  ({\bf 1}+{\cal Z}^{\dagger}{\cal Z})^{-1}{\cal Z}^{\dagger}        \\
  {\cal Z}({\bf 1}+{\cal Z}^{\dagger}{\cal Z})^{-1} & 
  {\cal Z}({\bf 1}+{\cal Z}^{\dagger}{\cal Z})^{-1}{\cal Z}^{\dagger}
  \end{array}
\right).
\end{equation}

\par \vspace{3mm}
Comparing (\ref{eq:oike-expression-2}) with (\ref{eq:quantum-projector}) 
we obtain the ``local coordinate" 
\begin{equation}
\label{eq:quantum local coordinate}
{\cal Z}=\frac{1}{R(N)+\theta}a^{\dagger}=a^{\dagger}\frac{1}{R(N+1)+\theta}
\end{equation}
where $R(N)=\sqrt{N+\theta^{2}}$. ${\cal Z}$ obtained by ``stereographic 
projection" is a kind of complex coordinate. 

Now if we take a classical limit $a \longrightarrow x-iy$, 
$a^{\dagger} \longrightarrow x+iy$ and $\theta=z$ then 
\begin{equation}
Z_{c}=\frac{x+iy}{r+z}
\end{equation}
where $r=\sqrt{x^{2}+y^{2}+z^{2}}$. This is nothing but a well--known one for 
(\ref{eq:projector}).

\vspace{5mm} \noindent
\begin{Large}
{\bf C\ \ Some Calculations of First Chern Class}
\end{Large}
\par \vspace{3mm} \noindent
We calculate the first Chern class of some vector bundles on $\fukuso P^{1}$ 
and show that the mapping degree of Veronese mapping is just $n$. 

We write our definition of $\fukuso P^{n}$ once more : 
\[
 {\fukuso}P^{n}= 
     \{P \in M(n+1; \fukuso)\ |\ P^{2} = P,\ P = P^{\dagger}\ \mbox{and}\ 
       \mbox{tr}P = 1 \}.
\]
On this space we define a canonical vector bundle like
\begin{eqnarray*}
E_{n}&=&\left\{(P,v) \in {\fukuso}P^{n} \times \fukuso^{n+1}\ |\ 
Pv=v\right\}, \\
\pi &:& E_{n}\ \longrightarrow\ {\fukuso}P^{n},\quad \pi(P,v)=P.
\end{eqnarray*}
Then the system $\xi_{n}=\{\fukuso, E_{n}, \pi, {\fukuso}P^{n}\}$ is called 
the canonical line bundle (because $P$ is rank one), see \cite{MN}, 
\cite{Fujii}. 
This is one of most important vector bundles.

Let us calculate the first Chern class of $\xi_{1}$. 
For the local coordinate $z$ in section 6.1, $P$ can be written as 
\begin{equation}
\label{eq:local projector}
P(z)=\frac{1}{1+|z|^{2}}
\left(
  \begin{array}{cc}
    1 & \bar{z}  \\
    z & |z|^{2}
  \end{array}
\right),
\quad 
v(z)=\alpha
\left(
  \begin{array}{c}
    1 \\
    z 
  \end{array}
\right)
\ (\alpha \in \fukuso).
\end{equation}
Then the canonical connection ${\cal A}$ and its curvature ${\cal F}$
can be written as 
\begin{equation}
\label{eq:canonical-connection}
{\cal A}=\frac{\bar{z}}{1+|z|^{2}}dz,
\quad
{\cal F}=d{\cal A}=\frac{1}{(1+|z|^{2})^{2}}d\bar{z}\wedge dz.
\end{equation}

Let $\chi$ be the Veronese mapping in section 6.1 ($\chi :\ \fukuso P^{1}\ 
\longrightarrow\ \fukuso P^{n}$). then we can consider the pull--back bundle 
$\chi^{*}\xi_{n} =\{\fukuso,\chi^{*}(E_{n}),\pi,\fukuso P^{1}\}$ where 
\begin{eqnarray*}
\chi^{*}(E_{n})&=&\left\{(P,v) \in \fukuso P^{1}\times \fukuso^{n+1}\ |\ 
\chi(P)v=v\right\} \\
\pi &:& \chi^{*}(E_{n})\ \longrightarrow\ {\fukuso}P^{1},\quad \pi(P,v)=P.
\end{eqnarray*}
See the following picture.

\setlength{\unitlength}{1mm} 
\begin{center}
\begin{picture}(80,50)
\put(20,16){\vector(1,0){30}}
\multiput(22,40)(2.5,0){10}{\line(1,0){2}}
\put(47,40){\vector(1,0){2}}
\put(12,35){\vector(0,-1){15}}
\put(57,35){\vector(0,-1){15}}
\put(10,13){\makebox(6,6)[c]{${\fukuso}P^{1}$}}
\put(55,13){\makebox(6,6)[c]{${\fukuso}P^{n}$}}
\put(30,17){\makebox(6,6)[c]{$\chi$}}
\put(10,37){\makebox(6,6)[c]{$\chi^{*}(E_{n})$}}
\put(54,37){\makebox(6,6)[c]{$E_{n}$}}
\end{picture}
\end{center}
\vspace{-10mm}

Let us give a local description.  For $z$ in (\ref{eq:local projector}) 
\[
\chi(P(z))=
\frac{1}{(1+|z|^{2})^{n}}
\left(
  \begin{array}{cc}
    1       & \psi(z)^{\dagger}        \\
    \psi(z) & \psi(z)\psi(z)^{\dagger}
  \end{array}
\right),
\quad 
v(z)=\alpha
\left(
  \begin{array}{c}
    1      \\
   \psi(z) 
  \end{array}
\right)
\ (\alpha \in \fukuso)
\]
where $\psi(z)$ is the map defined in section 6.1
\[
\psi(z)=
\left(
  \begin{array}{c}
    \sqrt{{}_nC_1}z \\
    \vdots \\
    \sqrt{{}_nC_j}z^{j} \\
    \vdots \\
    \sqrt{{}_nC_{n-1}}z^{n-1} \\
    z^{n}
  \end{array}
\right)
\quad \Longrightarrow \quad 1+\psi(z)^{\dagger}\psi(z)=(1+|z|^{2})^{n}.
\]

Now the connection and curvature of the pull--backed bundle are given by
\begin{equation}
{\cal A}_{n}=(1+\psi(z)^{\dagger}\psi(z))^{-1}\psi(z)^{\dagger}d\psi(z),
\quad
{\cal F}_{n}=d{\cal A}_{n}.
\end{equation}
Let us calculate : it is easy to see
\begin{eqnarray*}
{\cal A}_{n}
&=&\frac{{}_nC_1+\cdots+j{}_nC_j |z|^{2(j-1)}+\cdots+
n{}_nC_n |z|^{2(n-1)}}{(1+|z|^{2})^{n}} \bar{z}dz        \\
&=&\frac{\frac{d}{d(|z|^{2})}\left({}_nC_1|z|^{2}+
+\cdots+{}_nC_j |z|^{2j}+\cdots+{}_nC_n |z|^{2n}\right)}{(1+|z|^{2})^{n}}
\bar{z}dz    \\
&=&
\frac{\frac{d}{d(|z|^{2})}\left((1+|z|^{2})^{n}-1\right)}{(1+|z|^{2})^{n}}
\bar{z}dz    \\
&=&\frac{n(1+|z|^{2})^{n-1}}{(1+|z|^{2})^{n}}\bar{z}dz    \\
&=&n\frac{\bar{z}}{1+|z|^{2}}dz \\
&=&n{\cal A},
\end{eqnarray*}
therefore
\[
{\cal F}_{n}=n{\cal F}=n\frac{1}{(1+|z|^{2})^{2}}d\bar{z}\wedge dz.
\]
As a result we have 
\begin{equation}
\label{eq:first-Chern}
\mbox{Ch}_{1}(\chi^{*}\xi_{n})
=\frac{1}{2\pi i}\int_{\fukuso}n\frac{1}{(1+|z|^{2})^{2}}d\bar{z}\wedge dz
=n.
\end{equation}

As to calculations of geometric objects like Chern classes or holonomies 
on quantum computation see for example \cite{Fujii} or \cite{Fujii-2}.

\vspace{5mm} \noindent
\begin{Large}
{\bf D\ \ Difficulty of Tensor Decomposition}
\end{Large}
\par \vspace{3mm} \noindent
We point out a difficulty in obtaining the formula (\ref{eq:non-spin-1}) 
or (\ref{eq:non-spin-3/2}) by decomposing tensor products of $V$. 

To obtain the formula (\ref{eq:spin-1}) there is another method which 
uses a decomposition of the tensor product $A\otimes A$. Let us introduce. 
For 
\[
A=
\left(
  \begin{array}{cc}
    \alpha & -\bar{\beta} \\
    \beta  &  \bar{\alpha}
  \end{array}
\right)\ \in SU(2)
\]
we have 
\[
A\otimes A
=
\left(
  \begin{array}{cccc}
  \alpha^{2} & -\alpha\bar{\beta} & -\alpha\bar{\beta} & \bar{\beta}^{2} \\
  \alpha\beta & |\alpha|^{2} & -|\beta|^{2} & -\bar{\alpha}\bar{\beta}   \\
  \alpha\beta & -|\beta|^{2} & |\alpha|^{2} & -\bar{\alpha}\bar{\beta}   \\
  \beta^{2} & \bar{\alpha}\beta & \bar{\alpha}\beta & \bar{\alpha}^{2}
  \end{array}
\right).
\]
For the matrix $T$ coming from the Clebsch--Gordan decomposition 
\[
T=
\left(
  \begin{array}{cccc}
   0 & 1 & 0 & 0                                     \\
   \frac{1}{\sqrt{2}}  & 0 & \frac{1}{\sqrt{2}} & 0  \\
   -\frac{1}{\sqrt{2}} & 0 & \frac{1}{\sqrt{2}} & 0  \\
   0 & 0 & 0 & 1
  \end{array}
\right)
\]
it is easy to see 
\begin{equation}
T^{\dagger}(A\otimes A)T
=
\left(
  \begin{array}{cccc}
  |\alpha|^{2}+|\beta|^{2} &   &   &                             \\
    & \alpha^{2} & -\sqrt{2}\alpha\bar{\beta} & \bar{\beta}^{2}  \\
    & \sqrt{2}\alpha\beta & |\alpha|^{2}-|\beta|^{2} & 
      -\sqrt{2}\bar{\alpha}\bar{\beta}                           \\
    & \beta^{2} & \sqrt{2}\bar{\alpha}\beta & \bar{\alpha}^{2}
  \end{array}
\right)
=
\left(
  \begin{array}{cc}
   1 &             \\
     & \phi_{1}(A)
  \end{array}
\right)
\end{equation}
where we have used $|\alpha|^{2}+|\beta|^{2}=1$. This means a well--known 
decomposition
\[
\frac{1}{2}\otimes \frac{1}{2}=0\oplus 1.
\]

Let us take an analogy. For 
\[
V=
\left(
  \begin{array}{cc}
    X_{0} & -Y_{0}^{\dagger} \\
    Y_{0} &  X_{-1}
  \end{array}
\right)
\]
we have 
\[
V\otimes V
=
\left(
  \begin{array}{cccc}
   X_{0}^{2} & -X_{0}Y_{0}^{\dagger} & -Y_{0}^{\dagger}X_{0} & 
   Y_{0}^{\dagger}Y_{0}^{\dagger} \\
   X_{0}Y_{0} & X_{0}X_{-1} & -Y_{0}^{\dagger}Y_{0} & 
   -Y_{0}^{\dagger}X_{-1} \\
   Y_{0}X_{0} & -Y_{0}Y_{0}^{\dagger} & X_{-1}X_{0} & 
   -X_{-1}Y_{0}^{\dagger} \\
   Y_{0}Y_{0} & Y_{0}X_{-1} & X_{-1}Y_{0} & X_{-1}^{2}
  \end{array}
\right).
\]
However, the analogy breaks down at this stage because of 
the non--commutativity 
\begin{equation}
T^{\dagger}(V\otimes V)T
\ne
\left(
  \begin{array}{cc}
   {\bf 1} &              \\
           & \Phi_{1}(V)
  \end{array}
\right)
\end{equation}
for (\ref{eq:non-spin-1}). We leave it to the readers. 
There is no (well--known) direct metnod to obtain $\Phi_{1}(V)$ at the 
current time.

\par \vspace{5mm}
Last, let us make a commemnt. 
For the matrix $T$ coming from the Clebsch--Gordan decomposition (see 
\cite{papers})
\[
T=
\left(
  \begin{array}{cccccccc}
    0 & 0 & 0 & 0 & 1 & 0 & 0 & 0 \\
    \frac{1}{\sqrt{2}} & 0 & \frac{1}{\sqrt{6}} & 0 & 0 & 
    \frac{1}{\sqrt{3}} & 0 & 0 \\
    -\frac{1}{\sqrt{2}} & 0 & \frac{1}{\sqrt{6}} & 0 & 0 & 
    \frac{1}{\sqrt{3}} & 0 & 0 \\ 
    0 & 0 & 0 & \frac{\sqrt{2}}{\sqrt{3}} & 0 & 0 & \frac{1}{\sqrt{3}} & 0 \\
    0 & 0 & -\frac{\sqrt{2}}{\sqrt{3}} & 0 & 0 & \frac{1}{\sqrt{3}} & 0 & 0 \\
    0 & \frac{1}{\sqrt{2}} & 0 & -\frac{1}{\sqrt{6}} & 0 & 0 & 
    \frac{1}{\sqrt{3}} & 0 \\
    0 & -\frac{1}{\sqrt{2}} & 0 & -\frac{1}{\sqrt{6}} & 0 & 0 & 
    \frac{1}{\sqrt{3}} & 0 \\
    0 & 0 & 0 & 0 & 0 & 0 & 0 & 1
  \end{array}
\right)
\]
it is not difficult to see 
\begin{eqnarray}
&&T^{\dagger}(A\otimes A\otimes A)T   \nonumber \\
=
&&
\left(
  \begin{array}{cccccccc}
  \alpha & -\bar{\beta}  &   &   &   &   &   &       \\
  \beta  &  \bar{\alpha} &   &   &   &   &   &       \\
    &   & \alpha & -\bar{\beta}  &   &   &   &       \\
    &   & \beta  &  \bar{\alpha} &   &   &   &       \\
    &   &   &   & \alpha^{3} & -\sqrt{3}\alpha^{2}\bar{\beta} & 
    \sqrt{3}\alpha\bar{\beta}^{2} & -\bar{\beta}^{3} \\
    &   &   &   & \sqrt{3}\alpha^{2}\beta & (|\alpha|^{2}-2|\beta|^{2})\alpha 
    & -(2|\alpha|^{2}-|\beta|^{2})\bar{\beta} & 
    \sqrt{3}\bar{\alpha}\bar{\beta}^{2}              \\
    &   &   &   & \sqrt{3}\alpha\beta^{2} & (2|\alpha|^{2}-|\beta|^{2})\beta 
    & (|\alpha|^{2}-2|\beta|^{2})\bar{\alpha} & 
    -\sqrt{3}\bar{\alpha}^{2}\bar{\beta}             \\
    &   &   &   & \beta^{3} & \sqrt{3}\bar{\alpha}\beta^{2} & 
    \sqrt{3}\bar{\alpha}^{2}\beta & \bar{\alpha}^{3}
  \end{array}
\right)           \nonumber \\
=
&&
\left(
  \begin{array}{ccc}
  \phi_{1/2}(A) &     &     \\
      & \phi_{1/2}(A) &     \\
      &    & \phi_{3/2}(A)
  \end{array}
\right).
\end{eqnarray}
This means a well--known decomposition
\[
\frac{1}{2}\otimes \frac{1}{2}\otimes \frac{1}{2}=
\left(0\oplus 1\right)\otimes \frac{1}{2}=
\left(0\otimes \frac{1}{2}\right)\oplus
\left(1\otimes \frac{1}{2}\right)=
\frac{1}{2}\oplus \frac{1}{2}\oplus \frac{3}{2}.
\]

\vspace{5mm} \noindent
\begin{Large}
{\bf E\ \ Calculation of Some Integrals}
\end{Large}
\par \vspace{3mm} \noindent
We show some integrals.

\par \noindent
(A)\ Compact case :
\begin{equation}
<f|g>=\frac{2(2j+1)}{2\pi}\int_{\fukuso}
\frac{d^{2}z}{(1+|z|^{2})^{2j+2}}f(z)\overline{g(z)}
=\sum_{k=0}^{2j}\frac{1}{{}_{2j}C_k}a_{k}\bar{b}_{k}
\end{equation}
for $f(z)=\sum_{k=0}^{2j}a_{k}z^{k}$ and 
$g(z)=\sum_{k=0}^{2j}b_{k}z^{k}$ in ${\cal H}_{J}$. 

This is reduced to the equation 
\[
\frac{2(2j+1)}{2\pi}\int_{\fukuso}
\frac{d^{2}z}{(1+|z|^{2})^{2j+2}}z^{k}\bar{z}^{l}
=\delta_{kl}\frac{1}{{}_{2j}C_k}.
\]

\par \noindent
If we use the change of variables 
\[
x=\sqrt{r}\mbox{cos}\theta,\quad y=\sqrt{r}\mbox{sin}\theta
\quad \Longrightarrow \quad
d^{2}z=dxdy=\frac{1}{2}drd\theta
\]
then by using integration by parts 
\begin{eqnarray*}
\mbox{Left hand side}
&=&\delta_{kl}(2j+1)\int_{0}^{\infty}
\frac{r^{k}}{(1+r)^{2j+2}}dr  \\
&=&\delta_{kl}(2j+1)\frac{k}{2j+1}\int_{0}^{\infty}
\frac{r^{k-1}}{(1+r)^{2j+1}}dr  \\
&=& \cdots  \\
&=&\delta_{kl}(2j+1)\frac{k}{2j+1}\frac{k-1}{2j}\cdots \frac{1}{2j-k+2}
\frac{1}{2j-k+1}  \\
&=&\delta_{kl}\frac{k!}{(2j)(2j-1)\cdots (2j-k+1)}  \\
&=&\delta_{kl}\frac{1}{{}_{2j}C_k}.
\end{eqnarray*}

\par \noindent
(B)\ Non--compact case :
\begin{equation}
<f|g>
=\frac{2(2j-1)}{2\pi}\int_{D} d^{2}z(1-|z|^{2})^{2j-2}f(z)\overline{g(z)}
=\sum_{n=0}^{\infty}\frac{n!}{(2j)_{n}}a_{n}\bar{b}_{n}
\end{equation}
for $f(z)=\sum_{n=0}^{\infty}a_{n}z^{n}$ and 
$g(z)=\sum_{n=0}^{\infty}b_{n}z^{n}$ in $H^{2}$. 

This is reduced to the equation 
\[
\frac{2(2j-1)}{2\pi}\int_{D}d^{2}z(1-|z|^{2})^{2j-2}z^{k}\bar{z}^{l}
=\delta_{kl}\frac{k!}{(2j)_{k}}.
\]

Similarly in the case of (A), we obtain
\begin{eqnarray*}
\mbox{Left hand side}
&=&\delta_{kl}(2j-1)\int_{0}^{1}(1-r)^{2j-2}r^{k}dr  \\
&=&\delta_{kl}(2j-1)\frac{k}{2j-1}\int_{0}^{1}(1-r)^{2j-1}r^{k-1}dr \\
&=& \cdots  \\
&=&\delta_{kl}(2j-1)\frac{k}{2j-1}\frac{k-1}{2j}\cdots \frac{1}{2j+k-2}
\frac{1}{2j+k-1}  \\
&=&\delta_{kl}\frac{k!}{(2j)(2j+1)\cdots (2j+k-1)}  \\
&=&\delta_{kl}\frac{k!}{(2j)_{k}}
\end{eqnarray*}
by using integration by parts.


\end{document}